\documentclass[lettersize,journal]{IEEEtran}
\usepackage{amsmath,amsfonts}
\usepackage{algorithmic}
\usepackage{algorithm}
\usepackage{array}
\usepackage[caption=false,font=normalsize,labelfont=sf,textfont=sf]{subfig}
\usepackage{textcomp}
\usepackage{stfloats}
\usepackage{url}
\usepackage{verbatim}
\usepackage{graphicx}
\usepackage{cite}
\IEEEoverridecommandlockouts
\usepackage{amsmath,amssymb,amsfonts}
\usepackage{algorithmic}
\usepackage{graphicx}
\usepackage{textcomp}
\usepackage{xcolor}
\usepackage{caption}
\usepackage{float}

\begin{document}

\title{Design of Silicon Photonic microring resonators\\ with complex waveguide cross-sections \\ and minimal non-linearity}

\author{Stefania Cucco\textsuperscript{1}, Marco Novarese\textsuperscript{1}, Sebastian Romero Garcia\textsuperscript{2}, Jock Bovington\textsuperscript{3}, Mariangela Gioannini\textsuperscript{1} \\

~\IEEEmembership{\textsuperscript{1} Department of Electronics and Telecommunications, Politecnico di Torino.  Torino, Italy \\
\textsuperscript{2} Cisco Optical GmbH, Nuremberg, Germany \\
\textsuperscript{3} Cisco Systems, San Jose, CA 95134, USA \\ ~stefania.cucco@polito.it marco.novarese@polito.it mariangela.gioannini@polito.it}

\thanks{This work was partially supported by the European Union under the Italian National Recovery and Resilience Plan (NRRP) of NextGenerationEU, partnership on "Telecommunications of the Future" (PE00000001 - program "RESTART").}}        

\markboth{Journal of \LaTeX\ Class Files,~Vol.~14, No.~8, August~2021}%
{Shell \MakeLowercase{\textit{et al.}}: A Sample Article Using IEEEtran.cls for IEEE Journals}


\maketitle

\begin{abstract}
We present a model and a rigorous method to calculate the transmission coefficient of silicon micro-rings with complex waveguide cross-section including non-linear effects and self-heating, with very short simulation times. The method is applied to the design of MRRs in the SISCAP platform with high Q and reduced non-linearity, namely due to two photon absorption and free carrier absorption. We demonstrate that the free carrier diffusion in rib waveguides and Shockely-Read-Hall recombination play a fundamental role in reducing the impact of non-linearities in the ring.
\end{abstract}


\section{Introduction}
\begin{figure}[!t]
\begin{center}
\includegraphics[width=0.75\linewidth]{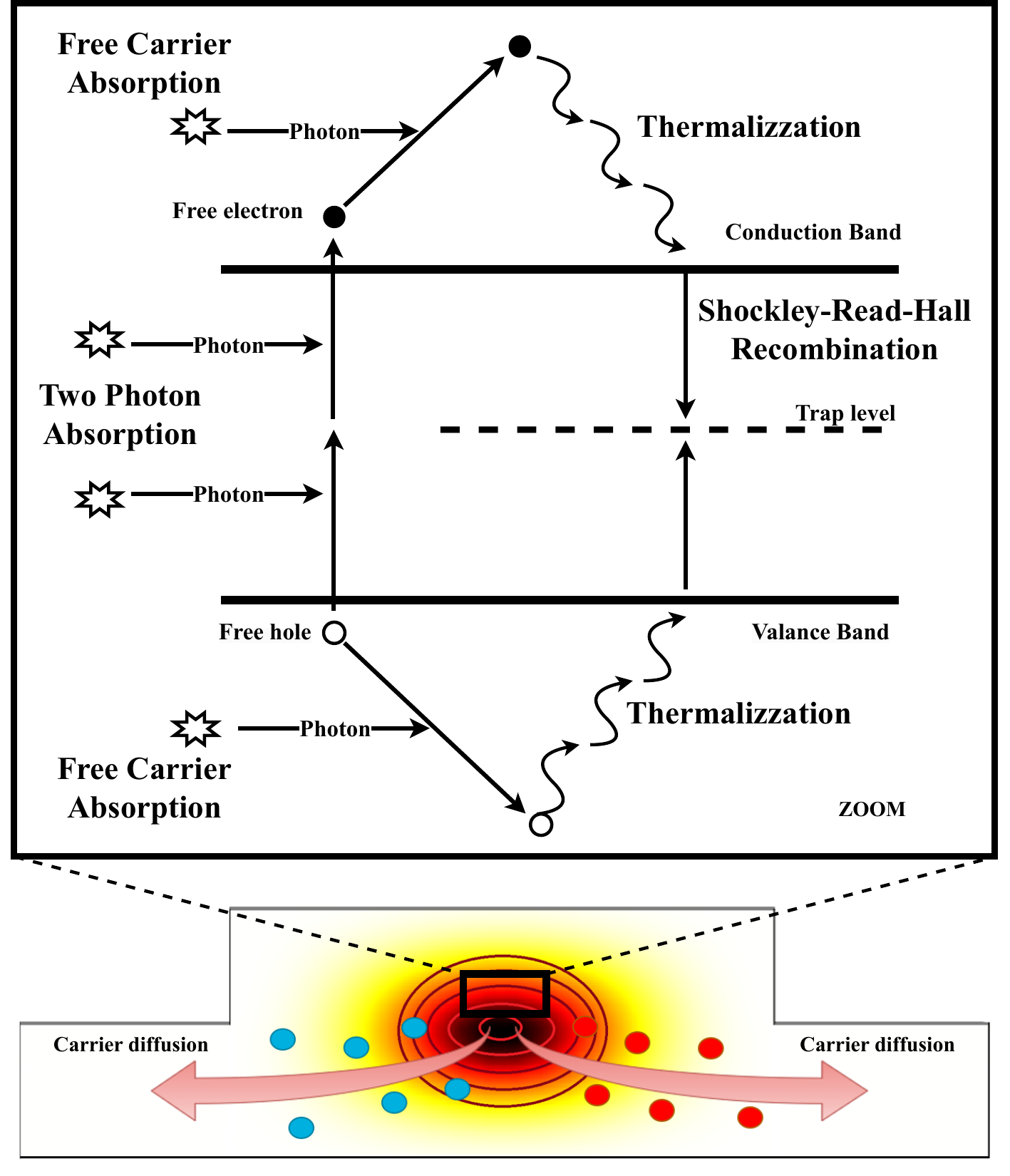} 
\end{center}
\caption{Sketch of the different nonlinear effects in silicon, involving transitions between the conduction and valence bands. }
\label{fig_effects}
\end{figure}

\IEEEPARstart{M}{icroring} resonators (MRRs) are useful in many silicon photonic integrated circuits. 
MRRs have a wide range of applications in photonics, including wavelength filtering, sensors, switching, and light modulation \cite{intro1,intro2}.  However, the non-linear response of silicon, even to moderate input powers, can limit the performances of high Q-factor MRRs \cite{intro3,intro3_bis}.
In this paper, we present  a  method to model and simulate  non-linear and thermal effects in silicon MRRs with any complex waveguide cross section; we apply it to the design of MRRs in the  SISCAP (semiconductor-insulator-semiconductor capacitor) platform \cite{c2}. This platform is beneficial for lowering the cost and energy consumption of optical transceivers for optical interconnects while maintaining an high data rate and volume in different applications \cite{intro4}. In the SISCAP platform, the silicon waveguide thickness is typically around 100 nm. Therefore, rib waveguides, which are commonly used to mitigate nonlinear effects in other, thicker platforms (e.g., 220 nm or 500 nm \cite{intro5}), cannot be utilized due to the high bend loss for radius smaller than $10\mu$\,m. Another approach for very high-Q and high finesse MRRs with reduced non-linear effects has been proposed recently in \cite{intro6}, but yet on thicker crystalline  silicon layer.\\
The main processes, summarized in Fig. \ref{fig_effects}, that cause nonlinear loss and dispersion in silicon are two-photon absorption (TPA) and free carrier absorption (FCA).  TPA occurs when two photons are absorbed, generating an electron-hole pair, while FCA occurs when the generated free carriers absorb another photon, promoting the electrons and the holes to higher energy in their bands. High energy free carriers (FCs) relax to lower energy and/or recombine with the valence band holes via  Shockely-Read-Hall (SRH)  and non-radiative recombination  releasing in both cases the energy in the form of heat (self-heating). This increase in temperature, along with nonlinear loss and dispersion, modifies the MRR transmission coefficient, causing a shift in the resonant wavelength and a reduction in the quality factor \cite{b1}.

For rib or more complex waveguide geometries (with cross-section that is not a simple strip) the generated FCs can diffuse in silicon  and heat is generated via SRH recombination following the carrier distribution as schematized for a rib cross section in Fig. \ref{fig_effects}. In this case, to evaluate  the degradation of the ring quality factor and the transmission spectrum, 2D simulations considering FCs generation via TPA, FC transport, non-radiative recombination, heat generation and dissipation are required. This self-consistent simulation set can provide the distribution of FCs and temperature allover the waveguide which allows to calculate the variation of modal loss and effective refractive index in the material. To the best of our knowledge, many other models simulating MRR including non-linear effects have been already published, but they always assume strip lumped waveguide and cannot be used to evaluate the impact of the waveguide geometry on the non-linear response of the MRR.  None of these previous works  includes the diffusion of carriers in rib waveguides and/or consider sessions of the waveguide having different materials ( such as for example silicon and poly-silicon) or dopings. The novelty of this work consists in presenting a rigorous self-consistent model for the calculation of the non-linear response of MRRs with a complex waveguide cross-sections. The methodology is applied for the simulation and design of non-linear MRRs that cannot be considered as simple lumped strip structures. While other techniques such as finite difference time domain (FDTD) simulations can be employed, they come with significant simulation time consumption. 
 For instance, we have performed a simulation of a strip waveguide ring of radius 3 µm using RSoft FullWave  and the FDTD method with the goal of extracting the ring transmission coefficient. We have chosen the FDTD model, since in RSoft it is the most appropriate method for including non-linear effects. For example, considering a dispersive material (ie: material refractive index that is wavelength dependent) and Kerr-non-linearity, this simulation lasts about 1 hour. Moreover, FCA cannot be easily included in the simulation with the models currently available in commercial tools. In contrast, using the same computer, our method utilizes Comsol Multiphysics simulations that require 2-15 minutes, 
depending on the complexity of the cross-section. The simulations include nonlinear effects (TPA and FCA) and self-heating, by varying the circular power by 12 values, using finite element method.

\begin{figure*}[!t]
  \centering
  \includegraphics[width=0.9\linewidth]{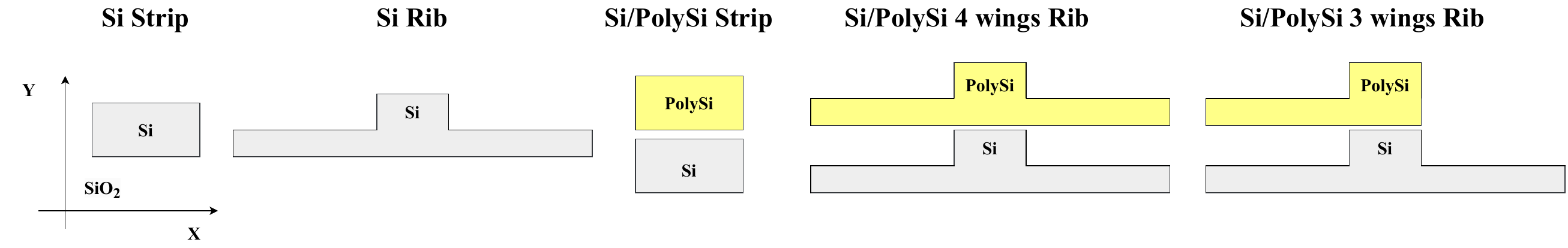}
\caption{Silicon and polysilicon waveguide cross sections  considered in this work according to the geometries available in the SISCAP platform.}
\label{fig1}
\end{figure*}

The paper is organized as follows: in Section II we present the numerical model to simulate FCs distribution and self-heating in the waveguide and we couple it to the simulation of the non-linear transmission coefficient of the MRRs. In section III we apply the method to the design of silicon waveguides and MRRs with small ring radius, high Q and minimal non-linear loss. In particular we consider 4  MRRs with different waveguide cross-sections, as shown in Fig. \ref{fig1}; they are in silicon and polysilicon according to the geometries available in the SISCAP platform. In Section IV, we validate the model by comparing measurements performed on an available microring resonator. Section V summarises the results of this work.

\section{Model}
In this section we introduce first the model to calculate the distribution of the free carriers in the waveguide cross-section accounting consistently for the self-heating; we then report the expression of the  variation of modal loss and effective refractive index and finally we compute the MRR transmission coefficient for any power incident in the MRR bus waveguide and for any input wavelength. 

\subsection{Modelling free carrier generation, transport and self-heating}
 We define with $P_c$, which is equal to the circulating power  in the ring,  the  optical power in a complex cross section of the ring.
 The TPA effect induces a carrier photo-generation rate $G_{ph}$ that it is defined per unit volume as \cite{b4}: 

\begin{equation}
   G_{ph}(x,y) = \dfrac{P_c^2}{\hbar\cdot \omega} \cdot  \beta_{TPA} \dfrac{n^2}{Z_0^2 \cdot 8 \cdot P_\mu^2}|e (x,y)|^4
\label{G_ph} 
\end{equation}
where $\hbar$ is the Plank's constant, $\omega$ is the angular frequency, $Z_0$ is the free-space wave impedance (377 $\Omega$)  and {\it n} is the refractive index of the material (silicon or polysilicon) where TPA occurs.  $\beta_{TPA}$ is TPA absorption coefficient  \cite{b1} and $P_\mu$ is the power normalization coefficient as defined in \cite{b4}. $\mathit{e(x,y)}$ (and $\mathit h(x,y)$ in the following) is the normalized optical electric (magnetic) field of the fundamental guided mode in the waveguide that, for the geometries we consider, is always assumed to be single mode. 
The optical mode profiles $\mathit{e(x,y)}$ and $\mathit{h(x,y)}$ are calculated with a finite element method and they are normalized with respect to the maximum peak of the electric field. Then, these profiles are imported in COMSOL Multiphysics where the transport model (COMSOL Semiconductor Module)  and the thermal model  (Heat Transfer Module) are coupled and solved self-consistently with a finite element method (FEM).
 The photogenerated  electrons in the conduction band  and holes in the valence band diffuse over the entire  silicon or polySi cross-sections as described by the drift-diffusion model \cite{drift-model}:

\begin{equation}
   \nabla^2 \Phi(x,y) = \dfrac{q}{\epsilon}(p(x,y)-n(x,y)-N_a^--f_t \cdot N_t)
\label{poisson} 
\end{equation}

\begin{align}
   \nabla(-D_n(x,y) \cdot \nabla n(x,y)+\mu_n(x,y) \cdot n(x,y) \cdot \nabla \Phi(x,y)) =&\nonumber\\
   = G_{ph}(x,y) - R_n(x,y)&
\label{drif_n} 
\end{align}

\begin{align}
   \nabla(-D_p(x,y) \cdot \nabla p(x,y)-\mu_p(x,y) \cdot p(x,y) \cdot \nabla \Phi(x,y)) =&\nonumber\\
   = G_{ph}(x,y) - R_p(x,y)&
\label{drif_p} 
\end{align}
Eq. (\ref{poisson}) is the Poisson's equation ruling the electrostatic potential ($\Phi$) as a function of the local charge density. In Eq. (\ref{poisson})  {\it q} is the  electric charge, $\epsilon$ is the electric permittivity of the material, $N_a^-$ is the silicon doping, {\it n(x,y)} and {\it p(x,y)}  are  the free electron and hole densities  and they are taken as the unknowns of our system of equations (\ref{poisson})-(\ref{Q}). $f_t$ is the occupancy  probability of an electron in the traps. $N_t$ is the bulk trap density per unit volume. We assume that the surface defects, denoted as $N_s$, are uniformly distributed across the entire interface between the silicon core and the silicon oxide. From this, we define an equivalent bulk trap density, $N_t$, which can be expressed as \cite{new}:
\begin{equation}
   N_t = N_s \cdot \dfrac{2(W+h)}{W \cdot h}
\label{Ns} 
\end{equation}

Eqs. (\ref{drif_n}-\ref{drif_p}) describe  the electrons and holes transport  at steady state; here the charge mobility and diffusion exhibit temperature-dependent behaviors, varying as functions of position $(x,y)$ according to the temperature distribution in the waveguide.\\ Considering the diffusion coefficient ($D_{n,p}$) we have: $D_{n,p}(x,y) = \mu_{n,p} \cdot k_B \cdot T(x,y)$, where $k_B$ is the Boltzmann's constant, {\it T(x,y)} is the material temperature in the cross-section and $\mu_{n,p}$ is the mobility. The temperature dependence of the mobility is modelled as \cite{b8}: 

\begin{equation}
   \mu_{n,p}(x,y) = \mu_{n0,p0} \bigg(\dfrac{T(x,y)}{T_{ref}}\bigg)^{-\alpha_n}
\label{mob} 
\end{equation}

where $\mu_{n0}$ and $\mu_{p0}$ are the mobilities of electrons and holes at the reference temperature ($T_{ref} = 300 K$) and $\alpha_n = 2.33$ is experimental exponential factor  for silicon \cite{b8}. At the RHS of Eqs. (\ref{drif_n}-\ref{drif_p})  $R_{n,p}(x,y)$ is the SRH non-radiative recombination rate dependent on the trap density and  trap energy level  position ($E_t$). 

\begin{figure}[!t]
\begin{center}
\includegraphics[width=0.9\linewidth]{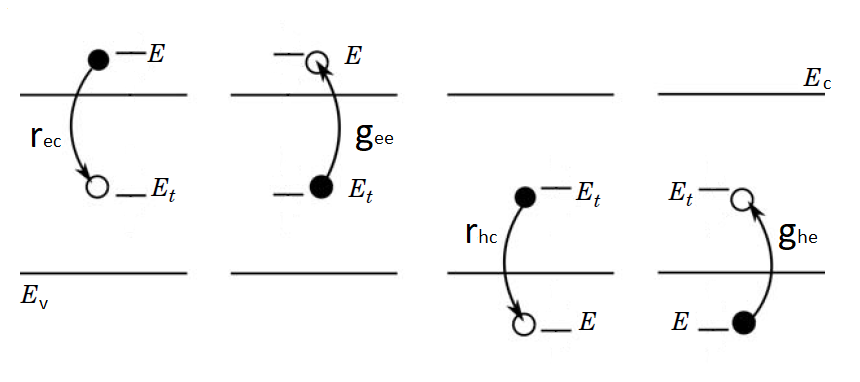} 
\end{center}
\caption{The SRH recombination consists of four processes: $r_{ec}$ is the capture of the electron from the conductive band to the trap level, and  $g_{ee}$ is the emission of the electron from the trap level to the conductive band. Similarly $r_{hc}$ and $g_{he}$ are respectively the capture and emission of a hole between the trap level and valence band.}
\label{fig_trap}
\end{figure}

We assume the trap energy level ($E_t$) to be positioned in the middle of the band-gap as in Fig. \ref{fig_trap}. 
$R_n$ ($R_p$ ) is calculated as the net  capture rate of  electrons  (holes) from the conduction band (valence band) to the trap state, namely  $R_n = r_{ec}-g_{ee}$ ( $R_p = r_{hc}-g_{he}$) . The set of Eqs. (\ref{poisson}-\ref{drif_p}) is implemented in the COMSOL Semiconductor Module \cite{b9}.  $r_{ec}$, $g_{ee}$, $r_{hc}$ and $g_{he}$ are defined  in Fig. \ref{fig_trap} as based on  \cite{b9};  they depend on the trap capture cross-section,  the trap density, and the occupancy of the trap by electrons ($f_t$) determined by the Fermi-Dirac statistics. At steady state $R_n$ is equal to $R_p$  and  can be expressed as \cite{b9,b10,b10bis}:

\begin{equation}
   R_n(x,y) = C_n \cdot N_t \cdot n(x,y) (1-f_t)\bigg(1-e^{\dfrac{E_{ft}-E_{fn}}{k_bT}}\bigg)
\label{Rn} 
\end{equation}

\begin{equation}
   R_p(x,y) = C_p \cdot N_t \cdot p(x,y) f_t \bigg(1-e^{\dfrac{E_{fp}-E_{ft}}{k_bT}}\bigg)
\label{Rp} 
\end{equation}

\begin{equation}
   f_t = \dfrac{1}{1+g_D \cdot e^{\dfrac{E_{t}-E_{ft}}{k_bT}}}
\label{ft} 
\end{equation}
where $C_{n,p}$ is the probability of an electron/hole capture  $ C_{n,p}= <\sigma_{n,p}> \cdot v^{th}_{n,p}$; $<\sigma_{n,p}>$ is the average capture cross section for electrons/holes  and $v^{th}_{n,p}$ is electron/hole thermal velocity. $g_D$ is the degeneracy factor of the level and it is equal to 1. $E_{fn}$ and $E_{fp}$ are the quasi-Fermi level of the free electrons (holes) in conduction (valance) band, while $E_{ft}$ is the quasi-Fermi level of the electron density in the trap\cite{b10bis}. 

Thermalization and Shockley-Read-Hall (SRH) recombination processes of free electron-hole pairs lead to an energy loss, resulting in the generation of heat and a  localized increase in temperature.
In this case we consider two factors that influence the temperature variation and heat dissipation: the propagation of heat within the ring structure for different radii, and the distance between the silicon waveguide and the silicon substrate.\\

The heat equation is expressed as:
\begin{equation}
   - \nabla ( d_z \cdot k_c\nabla T(x,y))= Q(x,y) \cdot d_z
\label{T} 
\end{equation}
$k_c$ is the material thermal conductivity and $d_z$ is the ring length calculated as $2 \pi R$, where $R$ is the  ring radius. Hence for a circular symmetry, the temperature profile is computed using the 2D axisymmetric formulation instead of the full 3D model, because the free carriers distributions do not change along z axis, that is perpendicular to the waveguide cross-section.   
$Q(x,y)$ is the heat source \cite{b11,b12}: 
\begin{equation}
   Q(x,y) = \dfrac{|J_n|^2}{\Sigma_n} + \dfrac{|J_p|^2}{\Sigma_p} +  Q_{FCA}(x,y) + R_{n} \cdot 2 \hbar \omega
\label{Q} 
\end{equation}
In eq. (\ref{Q}), the first two terms at RHS represent the heat generated by the Joule Effect.  $J_n$ and $J_p$ are the electron and hole current densities respectively, while $\Sigma_n$ and $\Sigma_p$ are the electrical conductivities  for electrons and holes;  they are defined as $\Sigma_n = q \cdot n(x,y) \cdot \mu_n$ and $\Sigma_p = q \cdot p(x,y) \cdot \mu_p$\cite{b5}. 

The third term at RHS is the heat generated by the thermalization of high energy FCs to the bottom of the conduction and valance bands (i.e., all light absorbed due to FCA is converted into heat). Thus $Q_{FCA}$ is the average optical power absorbed due to FCA per unit of volume in the MRR \cite{b1}:

\begin{equation}
   Q_{FCA}(x,y) = \rho_{abs}(x,y) \cdot I_{opt}(x,y)
\label{Qfca_eq} 
\end{equation}

 with  

\begin{equation}
   \rho_{abs}(x,y) = P_c (\alpha_0 + \Delta \alpha_{FC}(x,y)) (1+ t^2) 
\label{P_abs} 
\end{equation}

where $\alpha_0$ is the linear loss in the silicon microring,  $\Delta \alpha_{FC} (x,y)$  the local loss variation due to free-carrier absorption \cite{b3,b2}  and  $t$ is transmission coefficient at the MRR coupler, $t = \sqrt{1-k^2}$ in which $k$ is MRR coupling coefficient.  $I_{opt}$ accounts for the transverse profile of the guided mode:

\begin{equation}
   I_{opt}(x,y) =  \dfrac{\textit{Re} \{e(x,y) \, {\mathrm{x}} \,  h(x,y)^*\}\cdot \hat{z}}{2 \cdot P_\mu}
\label{Iopt_eq} 
\end{equation}

The last term at RHS of eq. (\ref{Q})  is  the heat generated by the non-radiative recombination: one electron-hole pair recombining via a trap releases the energy $2 \hbar \omega$ of the two-photons absorbed through TPA.

\subsection{Variation of modal loss and effective refractive index}
The calculation of the variation of modal loss and waveguide effective refractive index starts by computing the variation of the real (pedix $r$) and imaginary (pedix $i$) part of the permittivity \cite{b6,b7}: $\Delta \epsilon_{r,i} (x,y) = 2 \cdot \epsilon_0 \cdot n \cdot \Delta n_{r,i}(x,y) $ with $\epsilon_0$ being electrical vacuum permittivity;  $\Delta n_{r} (x,y)$  is the local variation of  refractive index caused by FCD and temperature and $\Delta n_i(x,y)$  the variation of the imaginary refractive index due to FCA. The variation of effective refractive index due to FCD is: 

\begin{equation}
    \Delta n_{eff,FCD} = \dfrac{  c \cdot \epsilon_0 \cdot n \int \int_\infty \Delta n_{FCD}(x,y) \cdot |e(x,y)|^2 dx dy}{\int \int_\infty \textit{Re} \{e(x,y) \, {\mathrm{x}} \, h(x,y)^*\} \cdot \hat{z} dx dy},
\label{nFCD_eq} 
\end{equation}

From this expression we can observe that, in the case of a constant carrier distribution over the whole cross-section (as in the case of a Si Strip waveguide), we obtain the  simple expression $\Delta n_{eff,FCD}=\Gamma \cdot \Delta n_{FCD}$ \cite{b3,b2}, where $\Gamma$ is optical confinement factor in the silicon core. $\Delta n_{FCD}$ is defined as \cite{b3}:

\begin{equation}
    \Delta n_{FCD}(x,y) = p_f \cdot n(x,y)^q + r_f \cdot p(x,y)^s
\label{Soref_n} 
\end{equation}

The  variation of refractive index due to self-heating is:
\begin{equation}
    \Delta n_ {eff,T} = \dfrac{c \cdot \epsilon_0 \cdot n \int \int_\infty \dfrac{dn}{dT} \cdot \Delta T(x,y) \cdot |e(x,y)|^2 dx dy}{\int \int_\infty \textit{Re} \{e(x,y) \, {\mathrm{x}} \, h(x,y)^*\} \cdot \hat{z} dx dy}
\label{nT_eq} 
\end{equation}
 
 where $\dfrac{dn}{dT} $ is the material thermo-optic coefficient (silicon or polySi).\\
 The optical modal loss due to FC is \cite{b7}:
\begin{equation}
    \Delta \alpha = \dfrac{2 c \cdot \epsilon_0 \cdot n \int \int_\infty \dfrac{\Delta\ \alpha_{FC}(x,y) \lambda}{4 \pi}\cdot |e(x,y)|^2 dx dy}{\int \int_\infty \textit{Re} \{e(x,y) \, {\mathrm{x}} \, h(x,y)^*\} \cdot \hat{z} dx dy}
\label{alpha_eq} 
\end{equation}

where $\Delta \alpha_{FC}(x,y)$ is defined as \cite{b3}:
\begin{equation}
    \Delta \alpha_{FC}(x,y) = a_f \cdot n(x,y)^b + c_f \cdot p(x,y)^d
\label{Soref_alpha} 
\end{equation}
the experimental parameters $a_f$, $b$, $c_f$, $d$, $p_f$, $q$, $r_f$ and $s$ in eqs. (\ref{Soref_n}) and (\ref{Soref_alpha}) depend on the material and they are wavelength dependent quantities ($\lambda = \mathrm{1.3\,\mu m}$ in our case); they are reported in Table \ref{tab:table1}.
The set of equations from eq. (\ref{G_ph}) to  eq. (\ref{Soref_alpha}) are solved together in COMSOL in order to get the modal loss and variation of the effective refractive index for a fixed circulating power $P_c$ in the ring.

\subsection{Modelling of the ring transmission coefficient}

The variation (as function of $P_c$) of modal loss and refractive index discussed in section B, can now be inserted in the non-linear model of the MRR to get the transmission coefficient for any bus input power and wavelength. The transmission coefficient of a MRR is defined as \cite{b1}:

\begin{equation}
    T_{thr} = \Big|\dfrac{t(1-k^2 \cdot a(P_c) \cdot e^{j\theta (P_c)})}{1-t^2\cdot a(P_c)\cdot e^{j\theta(P_c)}}\Big|^2
\label{T_thr} 
\end{equation}

where $a$ is propagation loss and $\theta$ is the phase variation of the field propagating in the MRR. Both terms depend non-linearly on the circulating power that it is expressed as:

\begin{equation}
    P_c = P_{bus} \cdot \dfrac{k^2(1-\eta^2)}{|1-t^2\cdot a(P_c)\cdot e^{j\theta(P_c)}|^2}
\label{P_c} 
\end{equation}
where $P_{bus}$ is the bus power inside the bus waveguides. The phase variation in eqs. (\ref{theta}) and (\ref{a_Pc})  is defined as:
\begin{equation}
    \theta(P_c) = \theta_0 + \dfrac{n_g}{c}(\omega - \omega_0)L+\Delta \theta(P_c)
\label{theta} 
\end{equation}
 $\theta_0$ is the phase variation referred to angular pulsation $\omega_0$ in linear regime in the waveguide, $n_g$ is group refractive index and L is the length of the ring. $\Delta \theta$ is the total phase variation per round trip due to non-linear effects and self-heating expressed as: 
\begin{equation}
    \Delta \theta(P_c) = \dfrac{2 \pi \cdot L}{\lambda_0} \cdot \Delta n_{eff}(P_c)
\label{Delta_theta} 
\end{equation}
$\Delta n_{eff}$ is the effective refractive index due to free carriers and temperature as a function of the circulation power.

The propagation loss of the optical field per round trip is
\begin{equation}
    a(P_c) = e^{\dfrac{-\alpha_{eff}(P_c)L}{2}}
\label{a_Pc} 
\end{equation}

where $\alpha_{eff}$ is the effective loss due to linear and non-linear effects and is defined as:

\begin{equation}
    \alpha_{eff}(P_c) = \alpha_0 + \alpha_{rad} + \Delta \alpha(P_c)
\label{alpha_eff} 
\end{equation}
 where $\alpha_0$ represents the linear loss component caused by light scattering, absorption at surface states, and minor residual doping, whereas $\alpha_{rad}$ corresponds to the bending loss, accounting for light that is radiated into the cladding.
By solving numerically the non-linear equation (\ref{P_c}), it is possible to find  $P_c$ for a fixed bus power and wavelength and then compute the MRR transmission coefficient with eq. (\ref{T_thr}).
 
\section{Results}

We consider five MRRs with different waveguide cross sections, as shown in Fig. \ref{fig1}: Si Strip, Si/PolySi Strip, Si Rib, Si/PolySi 4 wings Rib and Si/PolySi 3 wings Rib.  The material parameters of silicon and polySi used for the simulations are reported in Table \ref{tab:table1}, as extracted from simulations and experimental measurements from previous works. 

\begin{table}[!t]
  \begin{center}
    \caption{Parameters used in Comsol simulations for each waveguide cross-section.}
    \label{tab:table1}
    \begin{tabular}{|c||c||c||c|}
    \hline
      \textbf{Parameter} & \textbf{Silicon } & \textbf{Polysilicon } & \textbf{Unit}\\
      \hline
      $a_f$ & $8.8 \cdot 10^{-21}$ & $5.4 \cdot 10^{-20}$ & $cm^2$ \cite{b3,b2}\\
      $b$ & 1.167   &   1.167 & - \cite{b3,b2} \\
      $c_f$ & $5.84 \cdot 10^{-20}$ & $1.2 \cdot 10^{-19}$ & $cm^2$ \cite{b3,b2} \\
      $d$ & 1.109   &   1.109 & - \cite{b3,b2} \\
      $p_f$ & $5.4 \cdot 10^{-22}$ & $6.6 \cdot 10^{-22}$ & $cm^2$ \cite{b3,b2}\\
      $q$ & 1.011   &   1.011 & - \cite{b3,b2} \\
      $r_f$ & $1.53 \cdot 10^{-18}$ & $2.2 \cdot 10^{-18}$ & $cm^2$ \cite{b3,b2} \\
      $s$ & 0.838   &   0.838 & - \cite{b3,b2} \\
       $n$ & 3.45 & 3.45 & - \cite{b2}\\
      $\beta_{TPA}$ & $15 \cdot 10^{-12}$ & $15 \cdot 10^{-12}$ & $m/W$ \cite{b2}\\
      $\mu_{n0}$ & 1450  & 55.9 & $cm^2/(V \cdot s)$ \cite{b15,b2} \\
      $\mu_{p0}$ & 500  & 45  & $cm^2/(V \cdot s)$ \cite{b15,b2}\\
      $N_t$ & $5.454 \cdot 10^{16}$ & $5.36 \cdot 10^{17}$ & $1/cm^3$ \cite{b2}\\
      $<\sigma_n>$ & $2.18 \cdot 10^{-15}$ & $8.11 \cdot 10^{-16}$ & $cm^2$ \cite{b2}\\
      $<\sigma_p>$ & $1.09 \cdot 10^{-16}$ & $2.388 \cdot 10^{-15}$ & $cm^2$ \cite{b2}\\
      $v_n^{th}$ & $2.3 \cdot 10^{7}$ & $2.3 \cdot 10^{7}$ & $cm^2/s$ \cite{b15}\\
      $v_p^{th}$ & $1.65 \cdot 10^{7}$ & $1.65 \cdot 10^{7}$ & $cm^2/s$ \cite{b15}\\
      $\dfrac{dn}{dT}$ & $1.84 \cdot 10^{-4}$ & $1.84 \cdot 10^{-4}$ &$ K^{-1}$ \cite{b1}\\
      $\lambda$ & 1.3 & 1.3 & $\mu m$\\
       $k_c$ & 148 & 148 & $W/(m \cdot K)$ \cite{b14}\\
       $\alpha_0$ & 1.02 & 8 & $dB/cm$ \cite{b2}\\
      $N_a^-$ & $1\cdot 10^{15}$ & $1\cdot 10^{15}$ & $1/cm^3$ \\
     \hline
    \end{tabular}
  \end{center}
\end{table}

\begin{figure}[!t]
\begin{center}
\includegraphics[width=0.8\linewidth]{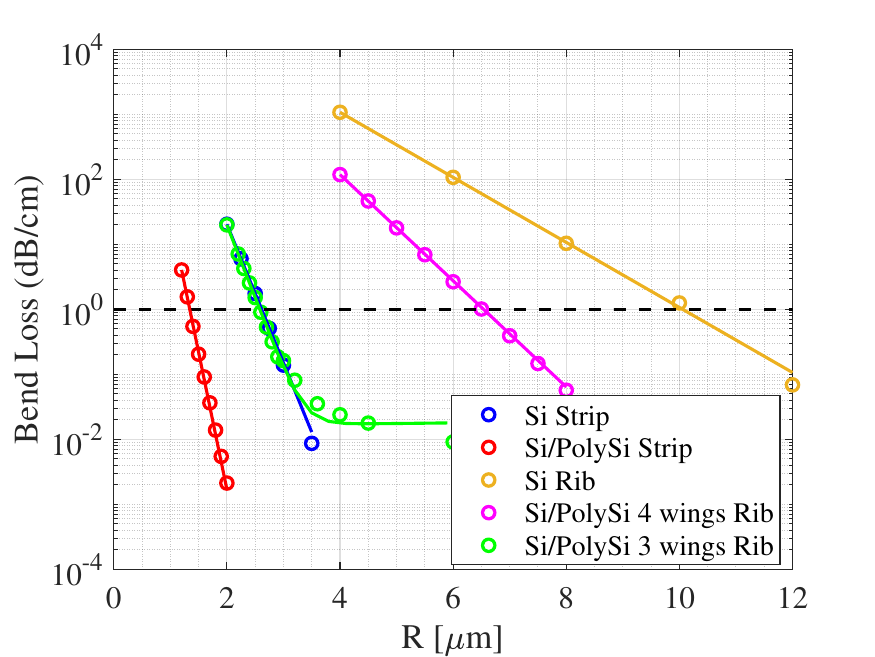} 
\end{center}
\caption{Bend losses as a function of radius for four type of MRRs with different cross-section. Exponential fits of the numerical results are reported with a solid line.}
\label{bend_loss}
\end{figure}

In Fig. \ref{bend_loss}, we report  the calculated bend loss for different waveguides versus bend radius. For each geometry, we determine the smallest allowed ring radius at which the bend loss exceeds a certain threshold $\mathrm{1\,dB/cm}$.

\begin{figure}[!t]
\begin{center}
\includegraphics[width=0.95\linewidth]{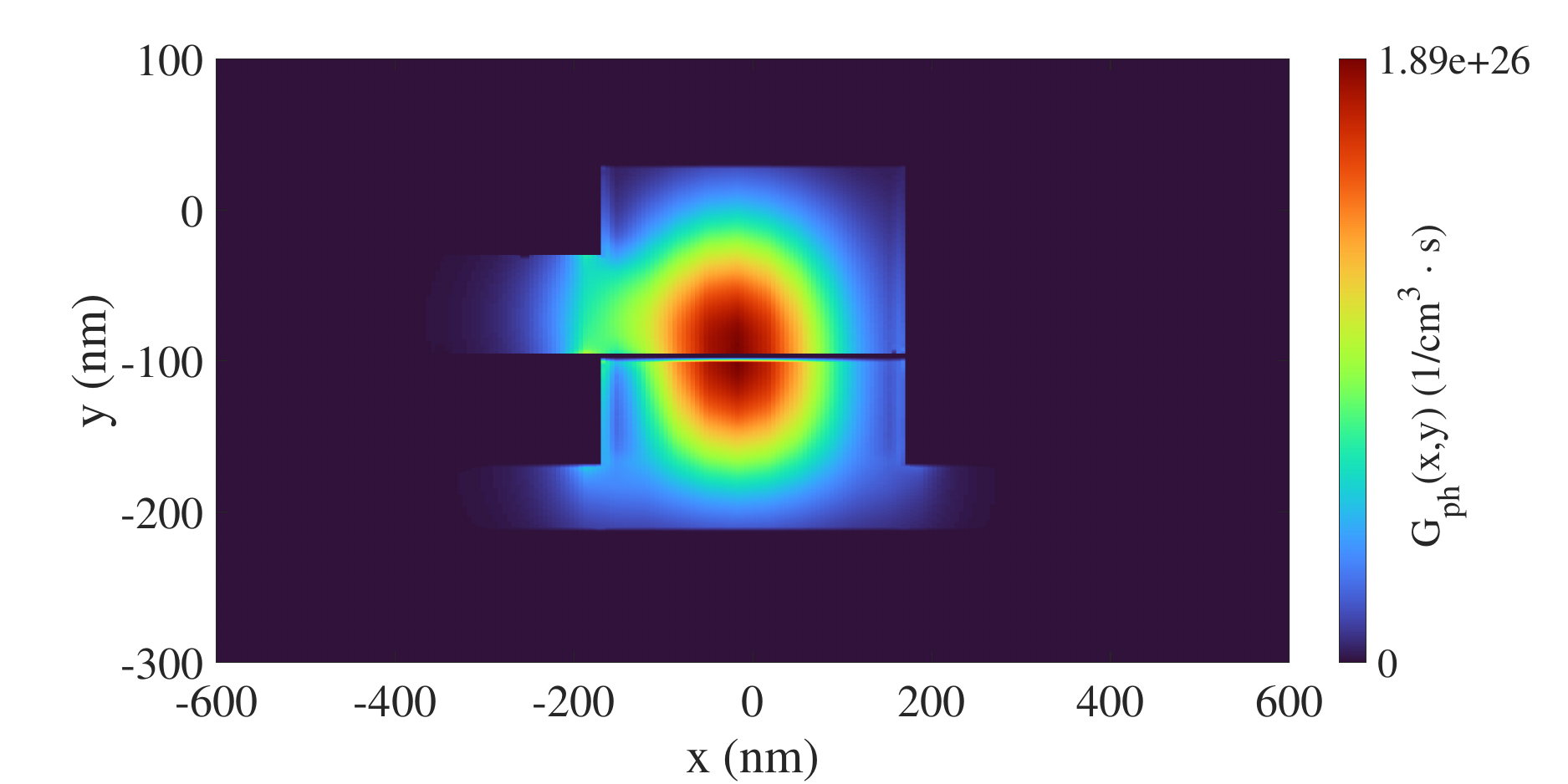} 
\end{center}
\caption{Example of FC generation rate due to TPA ($G_{ph}(x,y)$) with circulating power equal to $100$\,mW  in the case of the Si/PolySi 3 wings Rib waveguide.}
\label{Gph_2D}
\end{figure}

\begin{figure}[!t]
\begin{center}
\includegraphics[width=0.95\linewidth]{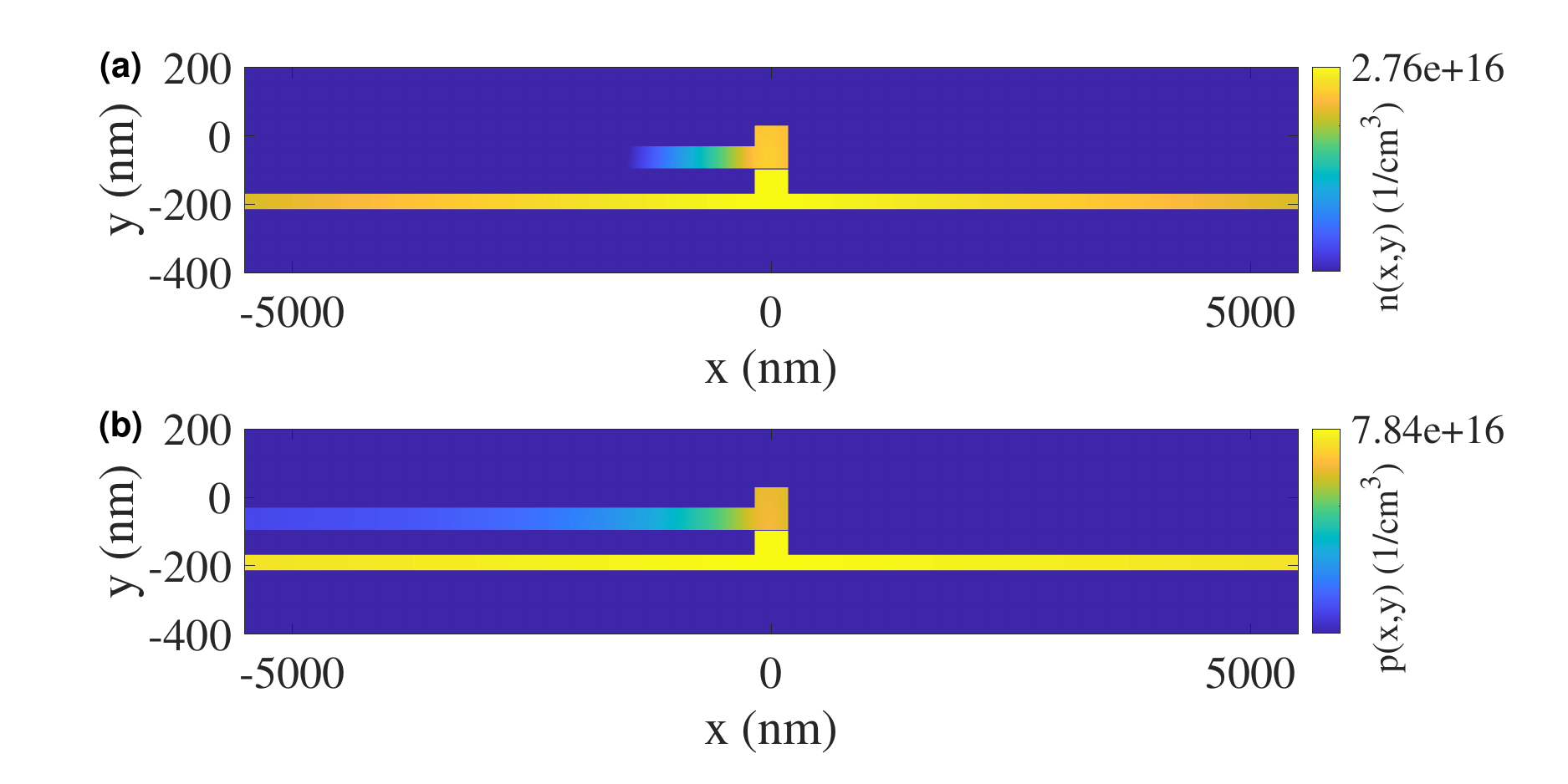} 
\end{center}
\caption{Example of carriers distributions: (a) electron distribution ($n(x,y)$) and (b) hole distribution ($p(x,y)$) in the Si/PolySi 3 wings Rib waveguide due to the photon-generation shown in Fig. \ref{Gph_2D}.}
\label{fig:elec-hole-distr}
\end{figure}

In this work we focus on  ring radii smaller than 6 $\mu m$  with  negligible bend losses ($<\mathrm{1\,dB/cm}$); the standard silicon rib structure and Si/PolySi 4 wings Rib do not meet these criteria because of the limited confinement of the optical field in the transversal direction. In this context, the addition of a polySi layer can enhance mode confinement, as observed in both the Si/PolySi strip and rib configurations. Therefore we exclude the Si rib from further consideration.\\
The radius for the three other rings has been fixed at $\mathrm{3.5\,\mu m}$ with a coupling coefficient $\mathrm{k^2} $ equal to $0.0055$ with  a resulting cold cavity quality factor  of $\mathrm{Q\approx30000}$.\\
An example of a COMSOL simulation of the Si/PolySi 3-wings rib waveguide is shown in Fig. \ref{Gph_2D}, which displays the FC generation rate with $P_c=100$\,mW. The carriers are photogenerated in the center of the waveguide and  then diffuse laterally in the wings as shown in Fig. \ref{fig:elec-hole-distr}. 

\begin{figure}[!t]
     \centering
     \subfloat[\label{tagli_3wing}]{\includegraphics[width=0.65\linewidth]{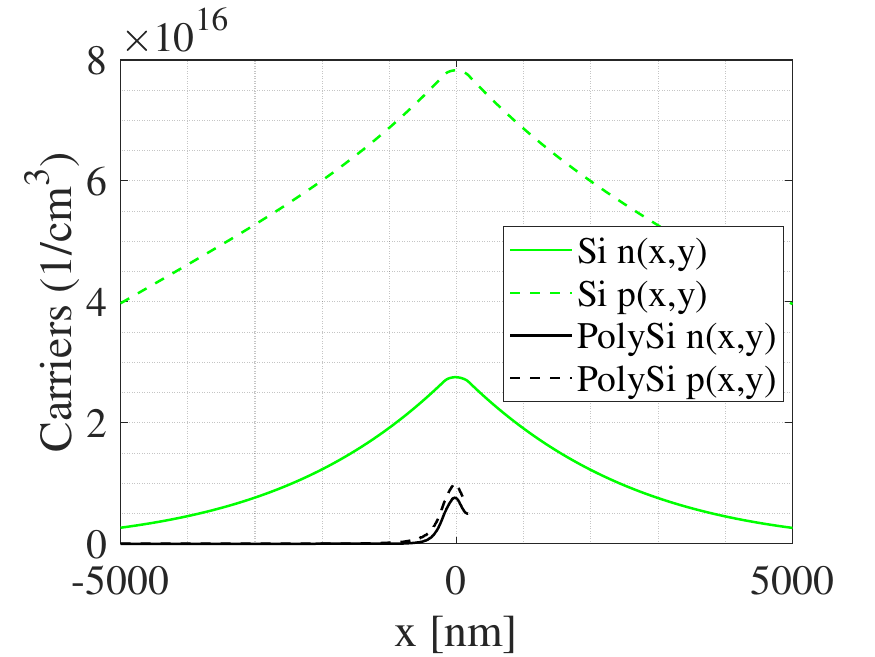} }
     \hfill
     \subfloat[\label{carriers_int}]{\includegraphics[width=0.65 \linewidth]{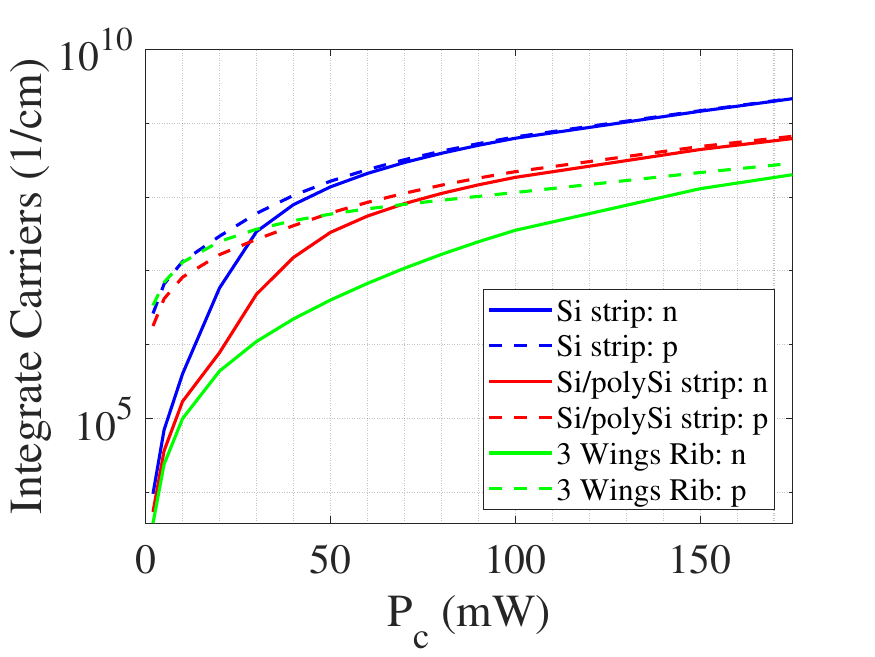}}
        
         \caption{(a) Distribution of carriers  along x-direction at $y=-62$\,nm and $y=-190$\,nm in the case of Fig.6. (b) Integrated carrier density as function of the circulating power for each waveguide analysed.}
        \label{fig:tagli}
\end{figure}

In Fig. \ref{fig:elec-hole-distr}  we note that the free carrier density is generally higher in the silicon cross section and carriers diffuse more with respect to the poly-Si case.
The smaller density of carriers in poly-Si can be explained by the higher SRH recombination rate caused by an higher number of defects in poly-Si (which  also reduces the diffusion length) as compared to silicon. Carrier distributions along x-direction at $y=-190$\,nm are also compared in Fig.\ref{tagli_3wing}.

Since FCA is proportional to the carrier density accumulated in the waveguide, we compare in Fig.7b the total number of  carriers  in the three different waveguides under the same circulating power $P_c$. The total number of carriers has been calculated by integrating  $n(x,y)$ and $p(x,y)$ over an equivalent area  defined as the $90 \%$ of the optical mode area.

\begin{figure}[!t]
     \begin{center}
     \subfloat[\label{Temp_si}]{\includegraphics[width=0.5\linewidth]{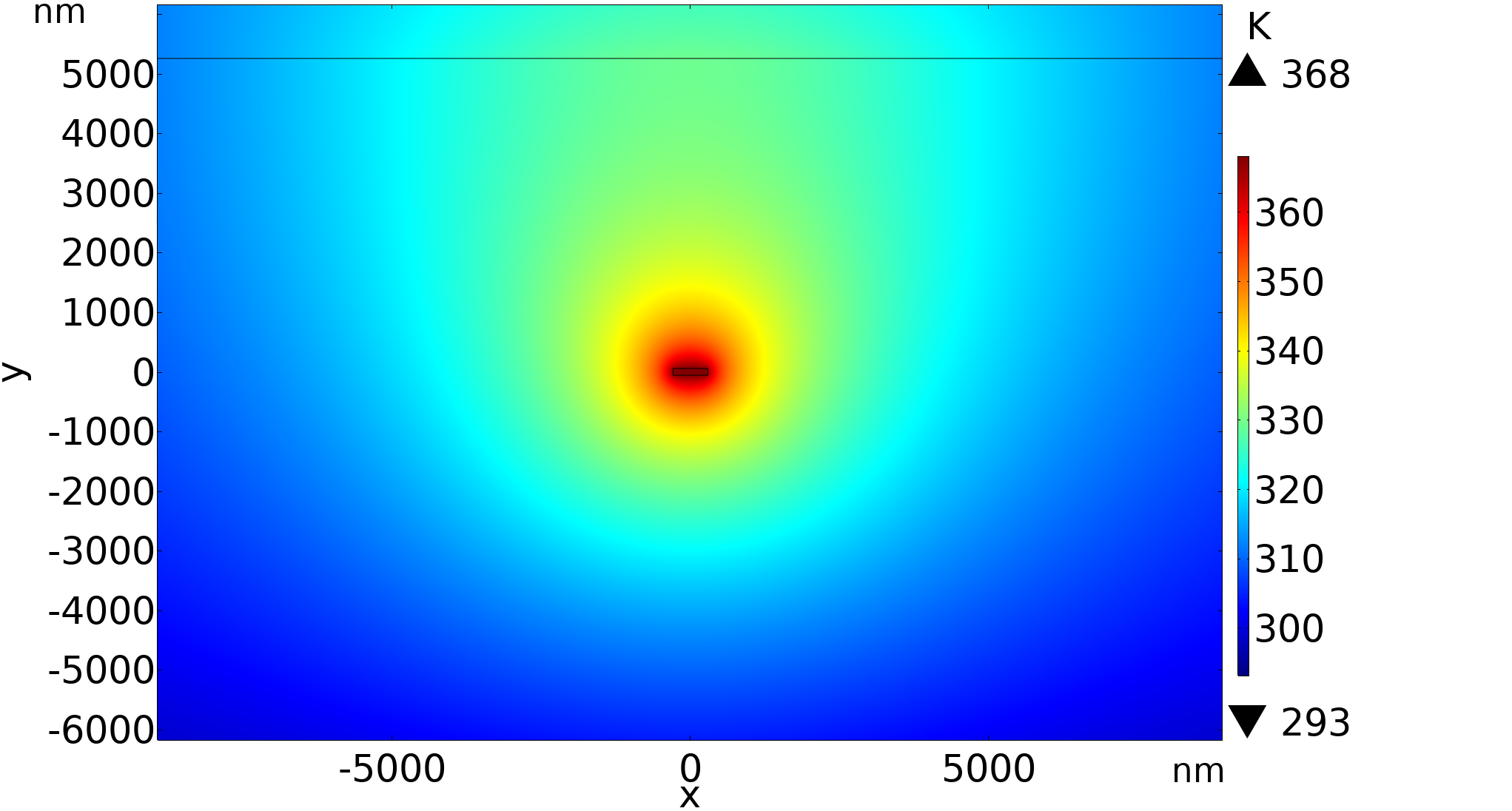}}
     \hfill
     \subfloat[\label{Temp_polysi}]{\includegraphics[width=0.5\linewidth]{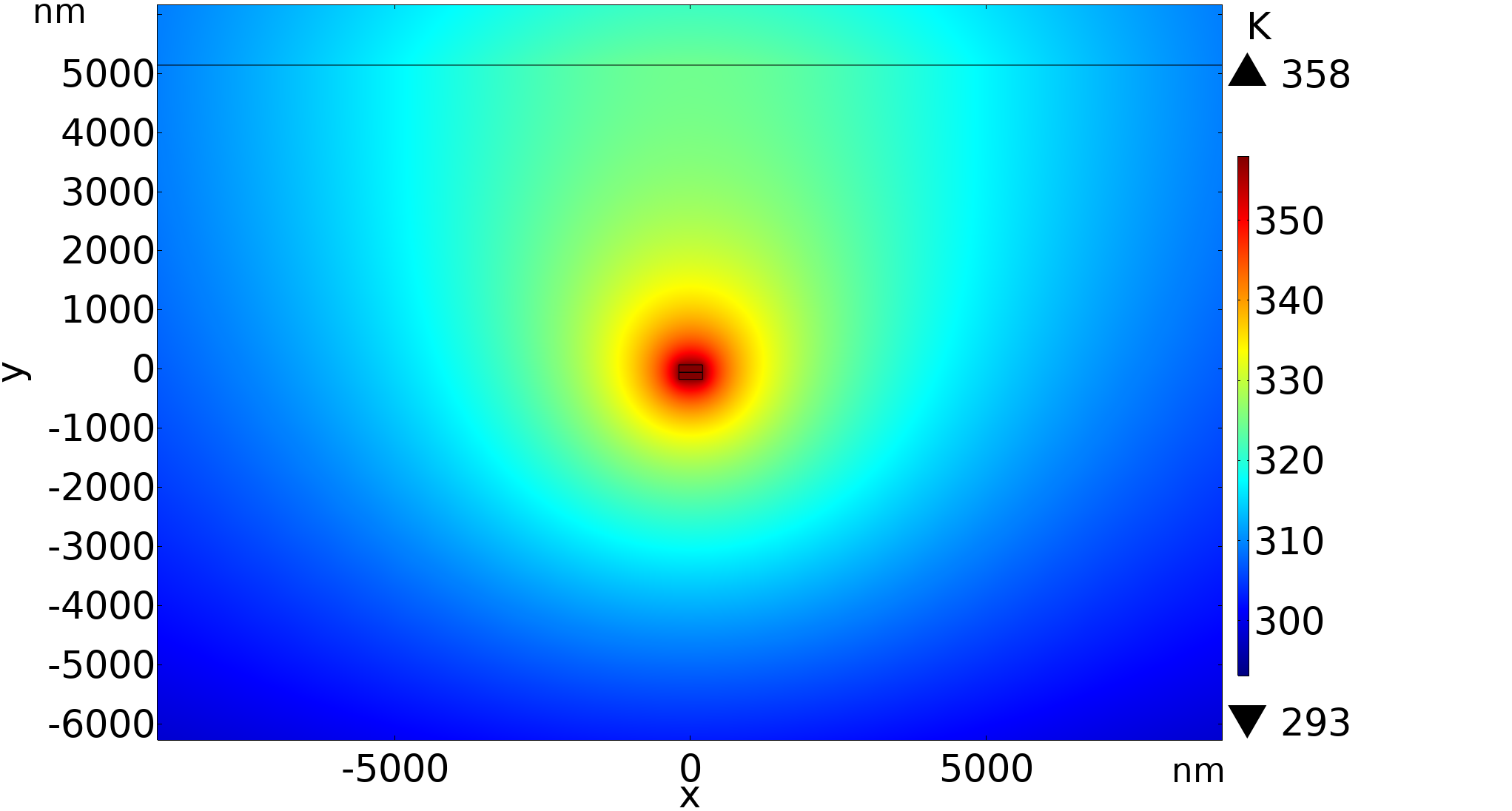}}
      \hfill
     \subfloat[\label{Temp_rib}]{\includegraphics[width=0.5\linewidth]{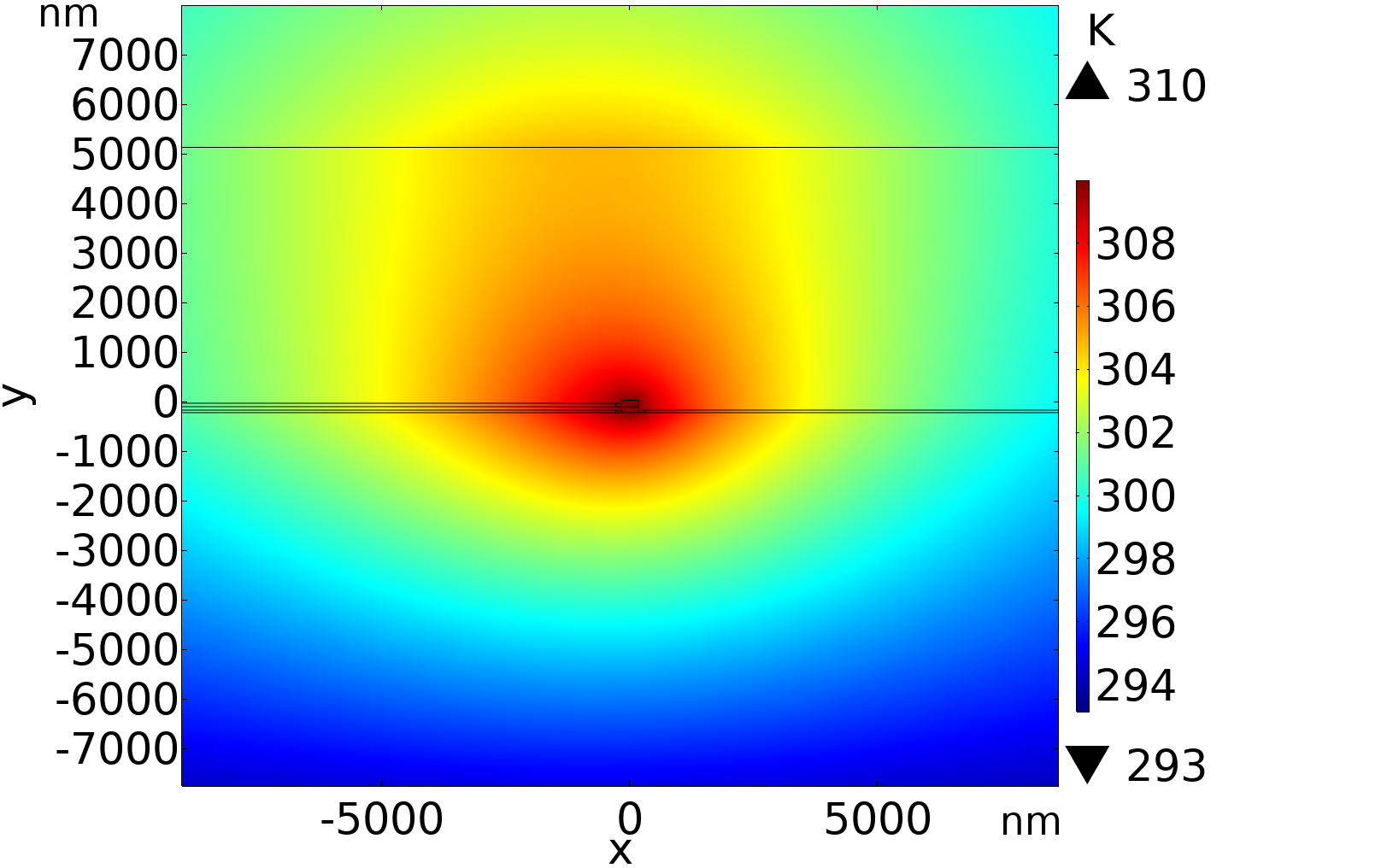}}
         \end{center} 
         \caption{Example of temperature variation due to self-heating for $P_c=100$\,mW in (a) Si Strip, (b) Si/PolySi Strip and (c) Si/PolySi 3 wings Rib.}
        \label{Temp2D}
         
\end{figure}

\begin{figure}[!t]
     \begin{center}
     \subfloat[ \label{Tmax_waveg}]
     {\includegraphics[width=0.5\linewidth]{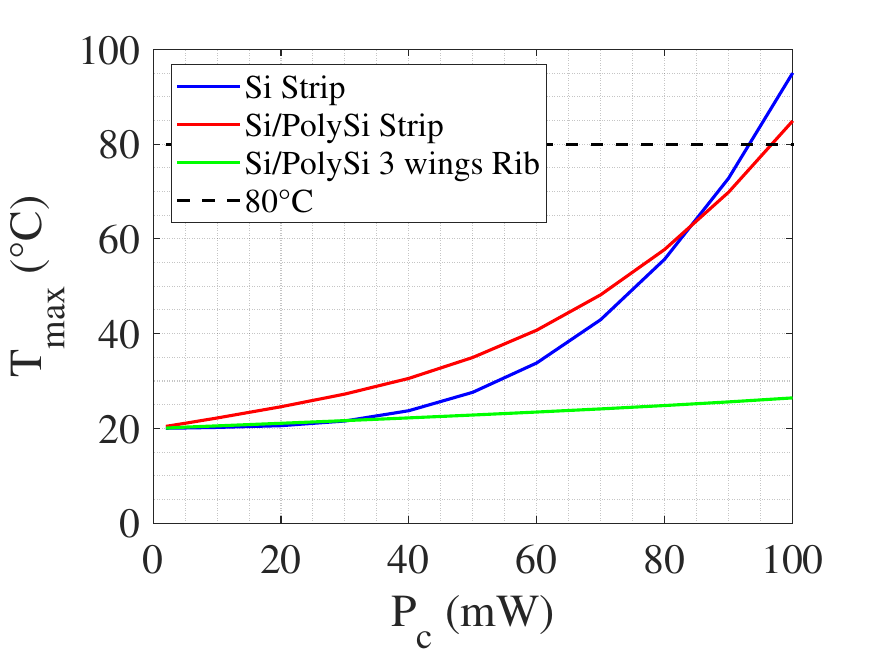}}
     \hfill
     \subfloat[\label{Qfca_int}]{\includegraphics[width=0.5\linewidth]{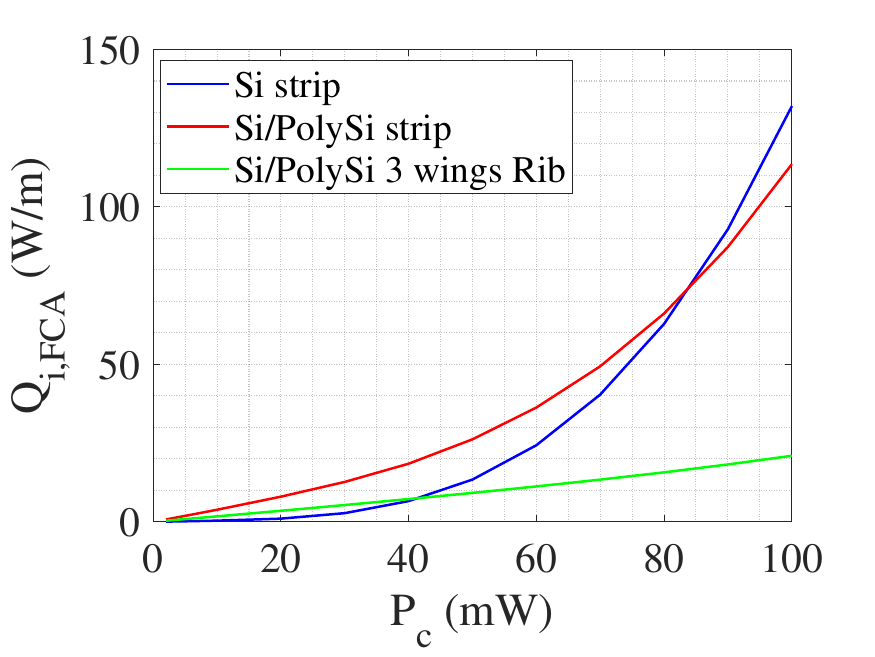}}
    
         \end{center} 
         \caption{(a) Maximum temperature of MRRs as a function of circulating power for each MRRs. (b) Heat source due to FCA integrated on the waveguide cross-section.}
        \label{Tem_rib}
         
\end{figure}

From a thermodynamic point of view, we have observed that the Si/PolySi 3 wings Rib waveguide heats up less than the other  waveguides operating at the same circulating power. This reduced heating is attributed to the lateral wings that facilitate heat dissipation, preventing it from being localised in the center of the waveguide. In Fig. \ref{Temp2D} we compare the temperature distribution and in Fig. \ref{Tmax_waveg} the maximum temperature reached in the core of the three waveguides. We have seen that in all cases, the major contribution to the temperature increase is due to the free carriers thermalization, as a result we compare in Fig. 9b the integrated $Q_{FCA}$ over the different waveguide cross-sections. The reduced number of carriers accumulated in the Si/PolySi 3 wings Rib reduces the FCA and therefore the carrier thermalization process too. 

Considering an operational temperature range for the device from $-40^o\,C$ to $80^o\,C$, the rib waveguide can efficiently handle circulating power levels of up to $200$\,mW at its maximum temperature due to heating. In contrast, other strip waveguides are limited to a maximum power of $100$\,mW, as illustrated  in Fig. \ref{Tmax_waveg}.

For circulating powers less than $80$\,mW, it is observed that the Si/PolySi Strip heats up more than the Si Strip, because the local modal losses of polysilicon are higher than in silicon.

\begin{figure}[!t]
     \centering
     \subfloat[ \label{alpha}]{\includegraphics[width=0.8\linewidth]{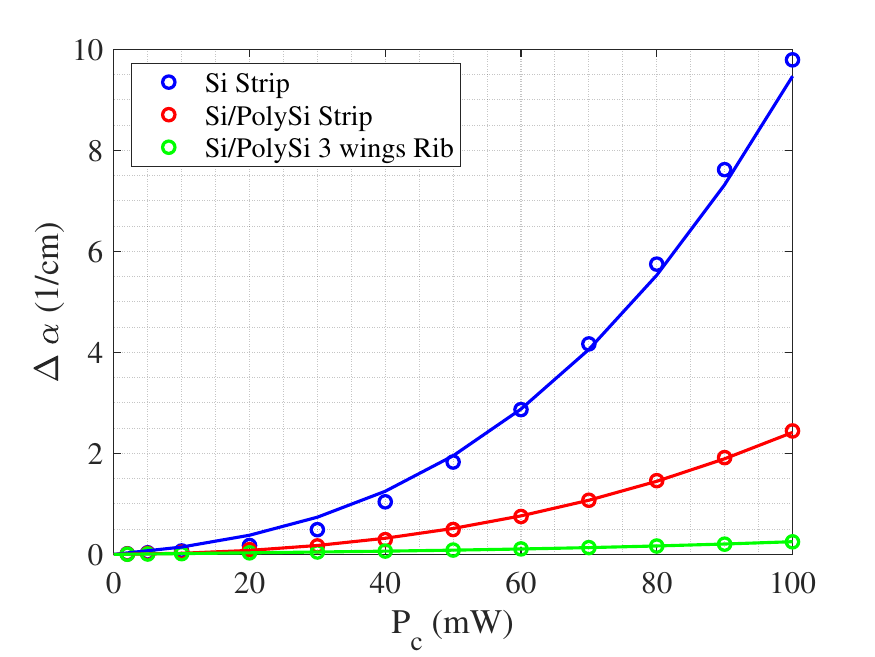}}
     \hfill
     \subfloat[\label{nFCD}]{\includegraphics[width=0.8\linewidth]{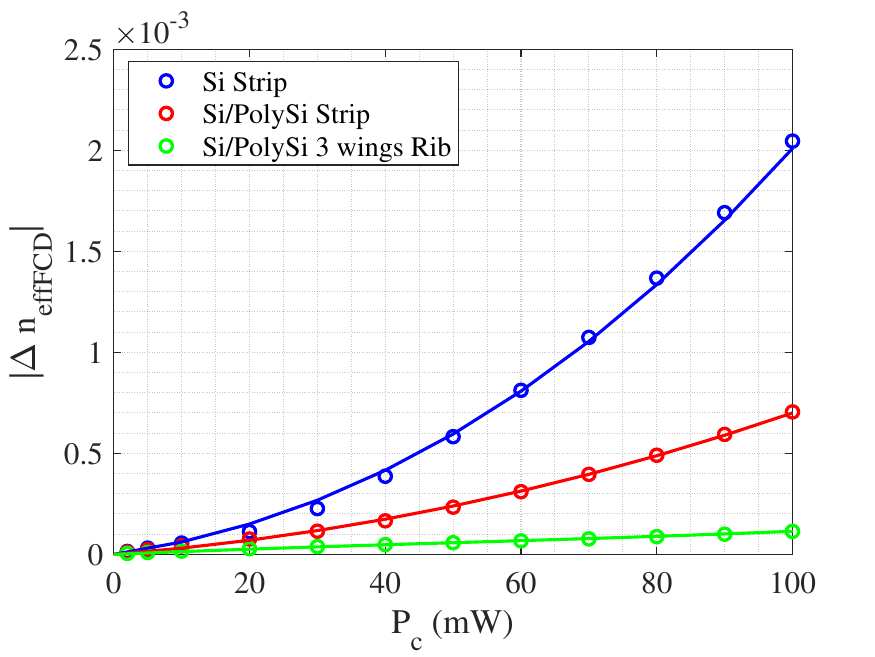}}
      \hfill
     \subfloat[\label{nT}]{\includegraphics[width=0.8\linewidth]{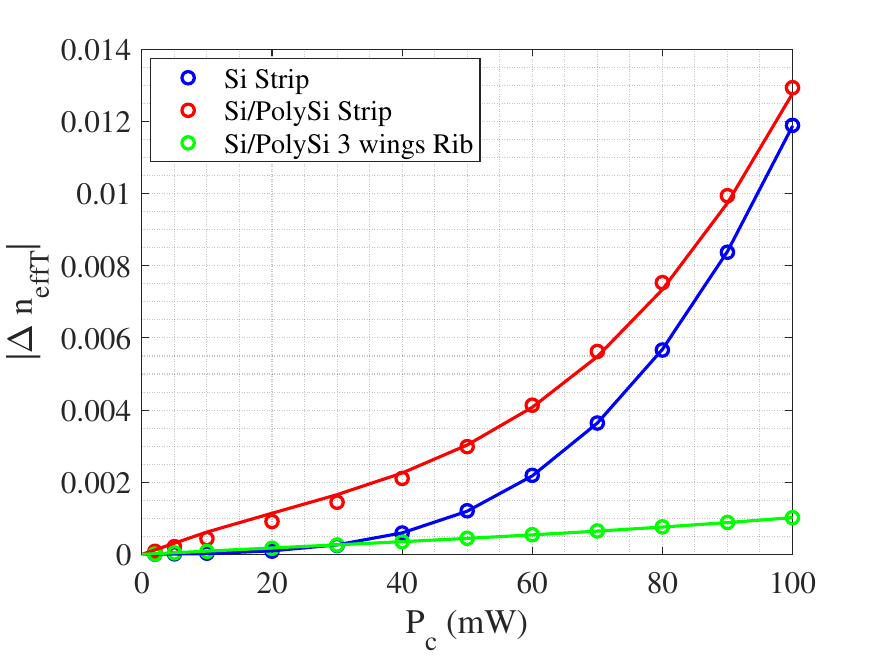}}
         
         \caption{Resulting variations of modal loss and effective refractive index as function of $P_c$ for each  MMR cross-section analyzed in this work. Solid lines represent  polynomial fits of the numerical results reported with symbols.}
        \label{fig:variation}
\end{figure}

In Fig. \ref{fig:variation}, we report the figure of merits of the three MRRs: $\Delta \alpha$, $\Delta n_{eff,FCD}$ and $\Delta n_{eff,T}$ as a function of the circulating power. Comparing in absolute value  $\Delta n_{eff,FCD}$ and $\Delta n_{eff,T}$, we observe that $\Delta n_{eff,T}$ is always greater than $\Delta n_{eff,FCD}$ in all waveguides, so we always expect a red shift of the transmission spectrum. As a consequence of the reduced carrier accumulation and self-heating discussed earlier, the Si/PolySi 3 wings Rib exhibits the lowest nonlinear effects for the same circulating power compared to the other two structures.

\begin{figure}[!t]
     \centering
     \subfloat[\label{T_si}]
     {\includegraphics[width=0.8\linewidth]{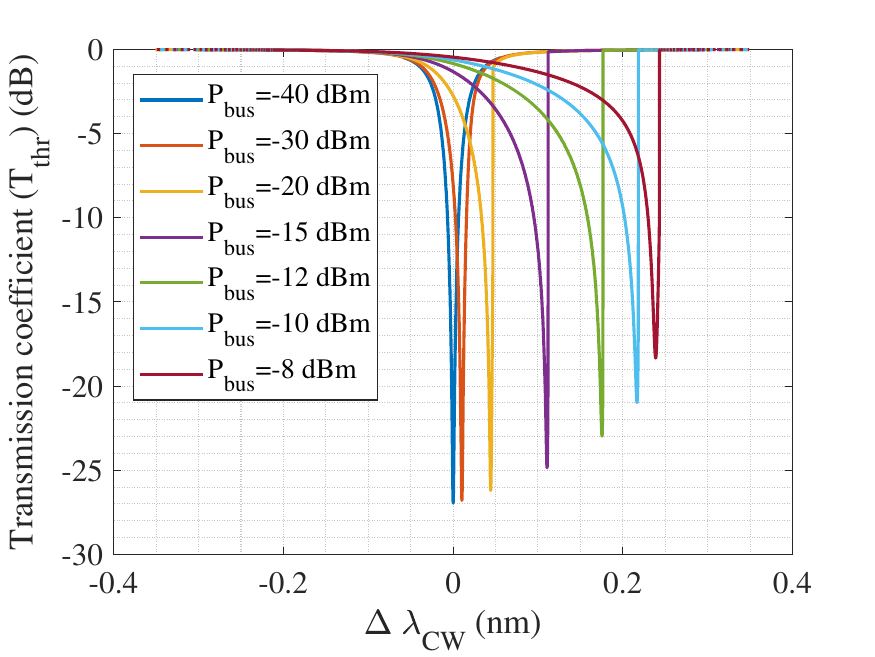}}        
     \hfill
     \subfloat[\label{T_polysi}]{\includegraphics[width=0.8\linewidth]{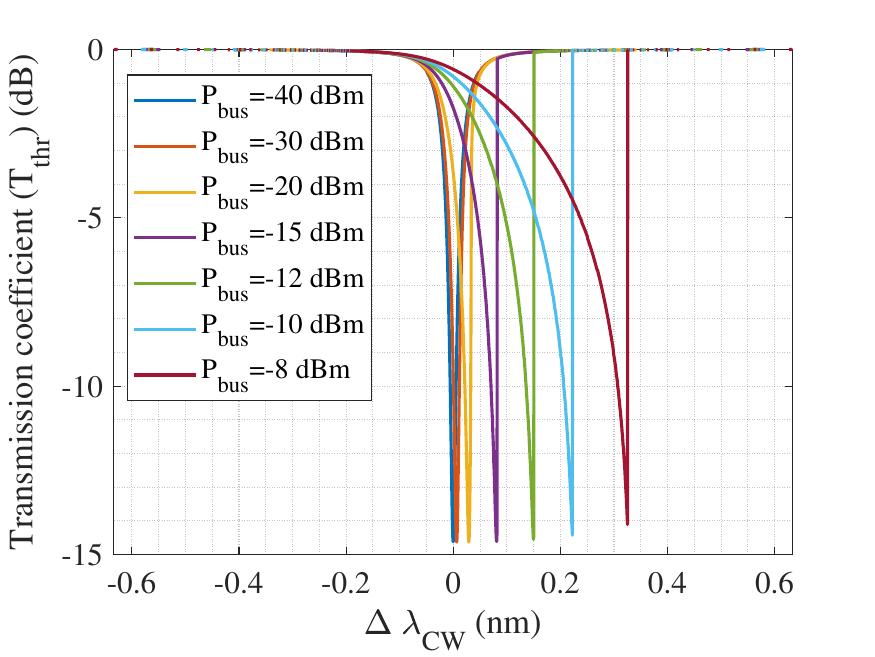}}     
      \hfill
     \subfloat[\label{T_3wings}]
     {\includegraphics[width=0.8\linewidth]{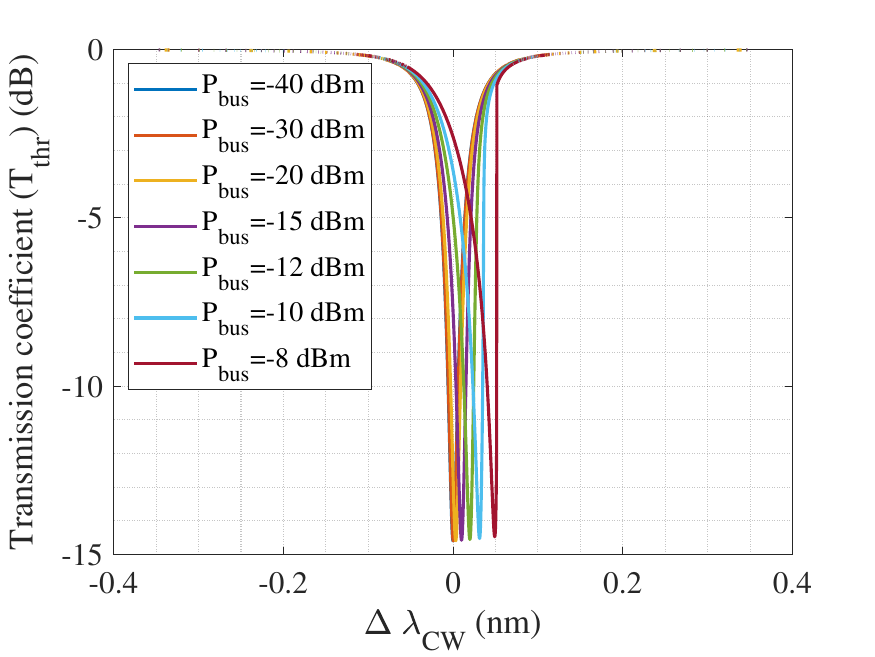}}
     
         \caption{ Transmission spectra (in dB) of: (a) Si Strip, (b) Si/PolySi Strip and (c) Si/PolySi 3 wings Rib at different bus input power with radius equal to $3.5\,\mu m$ and $k^2 = 0.0055$.}
        \label{fig:trasm}
\end{figure}

\begin{figure}[!t]
\begin{center}
\includegraphics[width=0.8\linewidth]{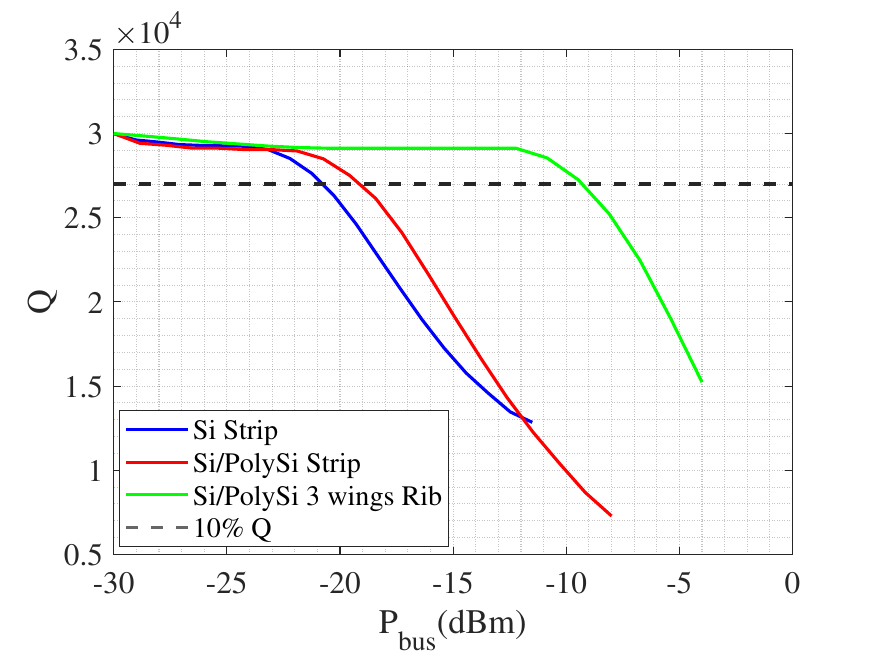} 
\end{center}
\caption{Degradation of the quality factor for the three structures as a function of the input power, with radius equal to $3.5 \mu m$ and $k^2 = 0.0055$.}
\label{Degrad_Q_new}
\end{figure}

The resulting variations of the figures of merits were fitted with third-degree polynomials as a function of the circulating power in the MRR. These polynomials are the input parameters for the model summarized in section II.C, that calculates the MRR transmission coefficient \cite{b1}. Fig. \ref{fig:trasm} reports respectively the MRR transmission spectrum of the Si Strip, Si/PolySi Strip and Si/PolySi 3 wings Rib. We can observe that the Si/PolySi 3-wings rib presents the smallest NL shift of the resonant wavelength among all the structures.\\

From the transmission coefficient, we can derive the degradation of the quality factor due to nonlinear effects as a function of the bus power\cite{b1} as shown in Fig. \ref{Degrad_Q_new}. The quality factor is defined as the central wavelength divided by the FWHM of the transmission coefficient and it is appropriate to carry out the calculation only within the stability regime (ie: when the ring is not oscillating or in the bi-stability region). It is evident that the Si/PolySi 3 wings Rib waveguide experiences less degradation in the  quality factor compared to the other two waveguides. Infact, with this waveguide cross-section, it is possible to achieve a maximum power level (with  $\mathrm{10\%}$  reduction of the Q) of  about $10$\,dB  higher than the standard Si Strip waveguide available on this platform.\\
The  free carrier lifetime is another important parameter that  quantifies how fast the generated free carrier are lost due to recombination process. 
To further compare the three waveguides, we proceed to calculate an equivalent free carrier lifetime. This calculation considers the rate at which free carriers recombine and diffuse out of the region where the optical field is confined. Considering free electrons, the equivalent free carrier lifetime is defined as:
\begin{equation}
    \tau_{eq,n} = \dfrac{ \int \int_{A_{eq}} n(x,y) dx dy}{\int \int_{A_{eq}} G_{ph}(x,y) dx dy}
\label{tau_eq} 
\end{equation}

\begin{figure}[!t]
\begin{center}
\includegraphics[width=0.8\linewidth]{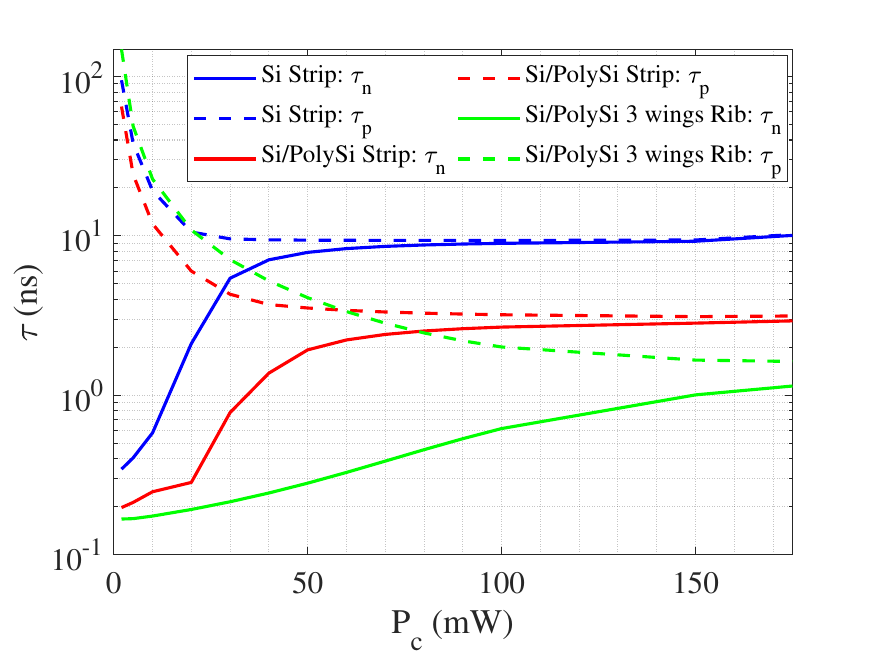} 
\end{center}
\caption{Equivalent electrons and holes lifetimes as a function of the circulation power. }
\label{tau}
\end{figure}

Where $A_{eq}$ is the equivalent area of the waveguide cross-section which contains the $90 \%$ of the confined optical mode. A similar expression can be written for the free holes by using the free hole density $p(x,y)$.\\
  The lifetimes of holes and electrons exhibit opposite trends. This can be explained by the fact that the capture cross section for holes is approximately one order of magnitude smaller than that for electrons \cite{new}. Consequently, at low circulating powers, the hole lifetime is significantly longer than the electron lifetime. As the power increases,  the carrier concentration rises, leading to a greater number of electrons being captured by the trap level. As a result, the electron lifetime increases with power because the trap level becomes saturated, causing electrons to remain there for a longer amount of time before recombining. At the same time, the hole lifetime decreases with increasing power since electrons released from the saturated trap level can recombine with holes, thus reducing their lifetime, as shown in Fig. \ref{tau}.
As it possible to see from Fig. \ref{tau}, the Si/PolySi 3 wings Rib has the shortest free carrier lifetime with respect to the other cases for high circulating powers.  This result arises from the free carriers diffusion and higher trap density associated with the polysilicon material. In all cases, the carrier lifetime depends on the circulating power, as we have already analyzed in our previous works \cite{b1,b2,marco_spie}.
The combination of these factors can explain the different lifetime trends observed for electrons and holes in various waveguide configurations.

\section{Comparison with experiments}
For the validation of the model, a Si/PolySi strip microring working at $1550$\,nm has been characterized in our laboratory. It has a  radius  $r=2\mu$\,m and coupling coefficients $k_1^2 = 0.08$ and $k_2^2= 0.09$ referring  to the coupling with the lower and upper bus waveguides \cite{b2}. 

We first simulated the structure in COMSOL and derived $\Delta \alpha$, $\Delta n_{eff,FCD}$ and $\Delta n_{eff,T}$ as a function of the circulating power. Then we used the MRR model developed in section II.C to calculate the actual circulating power in the ring and associated transmission spectra  for different bus power.

\begin{figure}[!t]
\begin{center}
\includegraphics[width=1\linewidth]{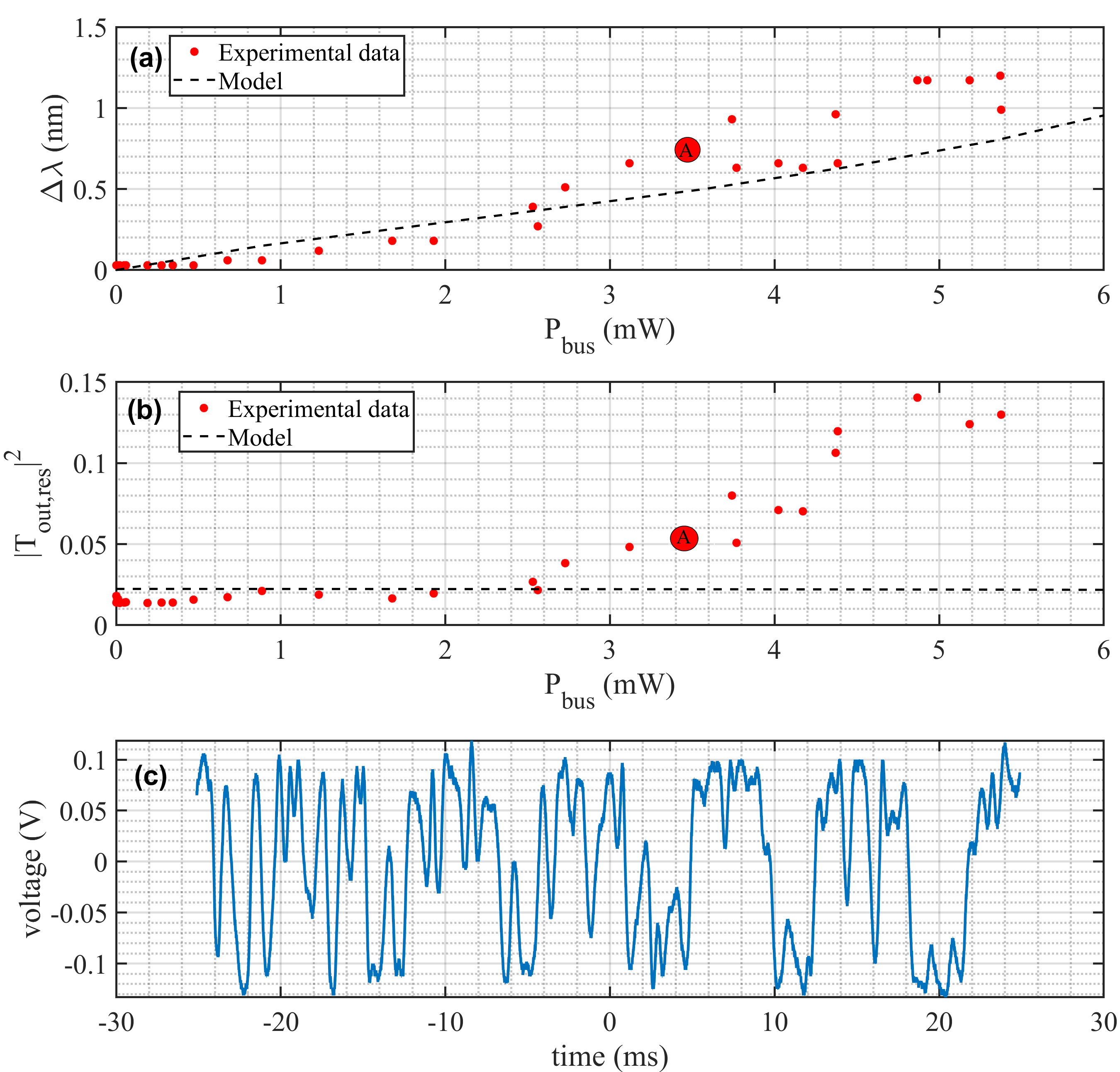} 
\end{center}
\caption{ Measured and modelled variation of the resonant wavelength (a) and transmission coefficients (b) for CW power for a MRR with a Si/PolySi Strip waveguide cross-section and radius equal to $2\,\mu$m. CW power fluctuations in time at the through port for a bus power equal to 3.5 mW (point A in Figure 14 (a) and (b) ).} 
\label{misure1}
\end{figure}

In order to measure the transmission coefficients of the ring in the non-linear regime, light from a tunable laser is coupled into the bus waveguide. The transmission at the through port is determined by taking the ratio of power detected at the through port out of resonance and the optical power in the bus waveguide. The transmission spectrum is obtained by gradually tuning the laser wavelength from shorter (blue) to longer (red) wavelengths across a specific cold ring resonance, denoted as $\lambda_0$. For this measurement, the ring was characterized using an Agilent 81980A tunable laser source with a wavelength sweep rate of 5 nm/s \cite{b1}.

Fig. \ref{misure1}(a) shows the measured and simulated variation of the ring resonant wavelength ($\Delta \lambda$) at the  through port as a function of $P_{bus}$ in the case of a forward wavelength sweep. 

In order to obtain a good fit in Fig \ref{misure1}, we assumed the following distribution of traps: in silicon the bulk trap density is $4.89 \cdot 10^{16} cm^{-3}$, whereas in polysilicon $N_t = 5.36 \cdot 10^{17} cm^{-3}$, and we considered a surface trap density in silicon of $2.18 \cdot 10^{11} cm^{-2}$ and in polysilicon $N_s = 2.59 \cdot 10^{12} cm^{-2}$. 

 Fig. \ref{misure1}(b) shows the measured and simulated transmission coefficients ($T_{out,res}$) at the resonance wavelength of Fig. \ref{misure1}(a) at the  through port as a function of $P_{bus}$. A good matching between simulation and measurements is obtained for input bus power up to 4 mW; the discrepancy at higher power is due to the fact that the measured MRR enters, for many input wavelengths,  in a self-oscillating regime where a transmission spectrum cannot be defined.  For injected wavelengths around the resonance and sufficient input power, the power measured at the through port oscillates over time, but the optical power meter reports an average value of this oscillating signal as shown in \ref{misure1}(c). In this scenario, the ring does not operate effectively, and the measured transmission cannot be considered a reliable metric for quantifying its static nonlinear response. As a result, the static model developed in this work is not applicable, and any comparison with the experimental transmission data would be misleading.

 \begin{figure*}[!t]
  \centering
  \includegraphics[width=0.9\linewidth]{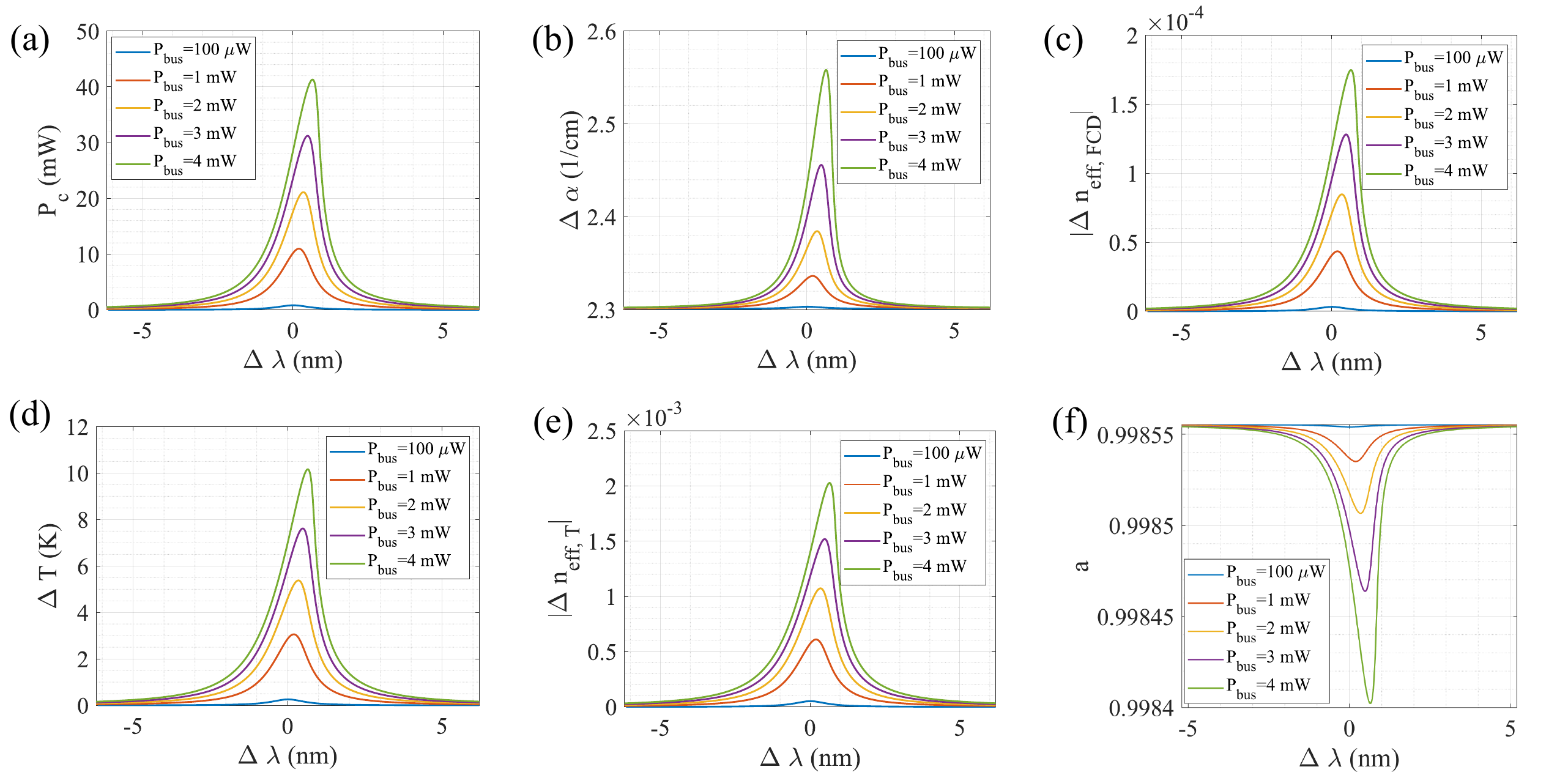}
\caption{ Influence of the bus power on some of the ring resonator figure of merits as a function fo the wavelength: (a) circulating power $P_c$, (b) variation modal losses $\Delta \alpha$, (c) refractive index shift caused by free carriers $\Delta n_{eff,FCD}$,  (d) temperature change due to self-heating $\Delta T$,  (e) refractive index shift induced by temperature $\Delta n_{eff,T}$, and (f) propagation loss $a$. }
\label{fig-merit}
\end{figure*}

As the input wavelength approaches the resonant wavelength, the circulating power increases (see Fig. \ref{fig-merit}(a)), leading to a rise in modal losses (Fig. \ref{fig-merit}(b)). A higher $P_c$ is also associated with an increase in free carrier generation, which reduces the effective refractive index due to free carrier dispersion (FCD), as shown in Fig. \ref{fig-merit}(c). This reduction causes a blue shift in the resonant wavelength.
The thermalization of free carriers releases energy in the form of heat, leading to a temperature increase within the ring (see Fig. \ref{fig-merit}(d)). This, in turn, increases the effective refractive index due to self-heating (Fig. \ref{fig-merit}(e)). It is important to note that the variation in the refractive index caused by self-heating is larger in magnitude than FCD, ultimately resulting in a red shift in the transmission spectrum.
  Fig. \ref{fig-merit} (f) demonstrates that the extinction ratio  of the transmission coefficient remains nearly  constant, while the resonance undergoes  a red shift,  as the bus power increases, since  the propagation losses are close to 1.
Fig. \ref{fig-merit} shows how the fundamental parameters
of the model vary as a function of the wavelength variation for
different bus powers up to 4 mW, because the deviations for
higher powers are due to the self-oscillation of the ring [9].



\section{Conclusion}
In summary, we have developed a model to compute the 2D distribution of free carriers and temperature in complex waveguide cross-sections of MRRs within the SISCAP platform, applying the FEM method to drastically reduce simulation times, as compared to other approaches such as FDTD. We stress that our approach propose a tool to simulate and design MRR with waveguides that can not be considered as simple strip waveguides.
 Importantly, our approach provides a tool to simulate and design MRRs featuring waveguides that cannot be approximated as simple strip waveguides. Traditional lumped models are typically used for the latter. Deriving lumped models for rib structures or other complex structures considered in this work is challenging due to their geometry. In such cases, carrier diffusion within the rib must be accounted for, which requires a model like the one we propose. 

We have successfully identified a design solution that minimizes non-linear effects and self-heating. The Si/PolySi 3 wings Rib waveguide stands out as the most favorable structure among those analyzed. The reason is that the rib waveguide facilitates free carrier diffusion and heat dissipation, while the addition of the poly-Si waveguide reduces bend loss and free-carrier lifetime by enhancing (SRH) recombination in poly-Si traps. This demonstrates the power and generality of the developed model for the understanding and design of high-Q MRRs with any geometry and material. \\




\vspace{11pt}

\vskip 0pt plus -1fil
\begin{IEEEbiography}[{\includegraphics[width=1in,height=1.25in,clip,keepaspectratio]{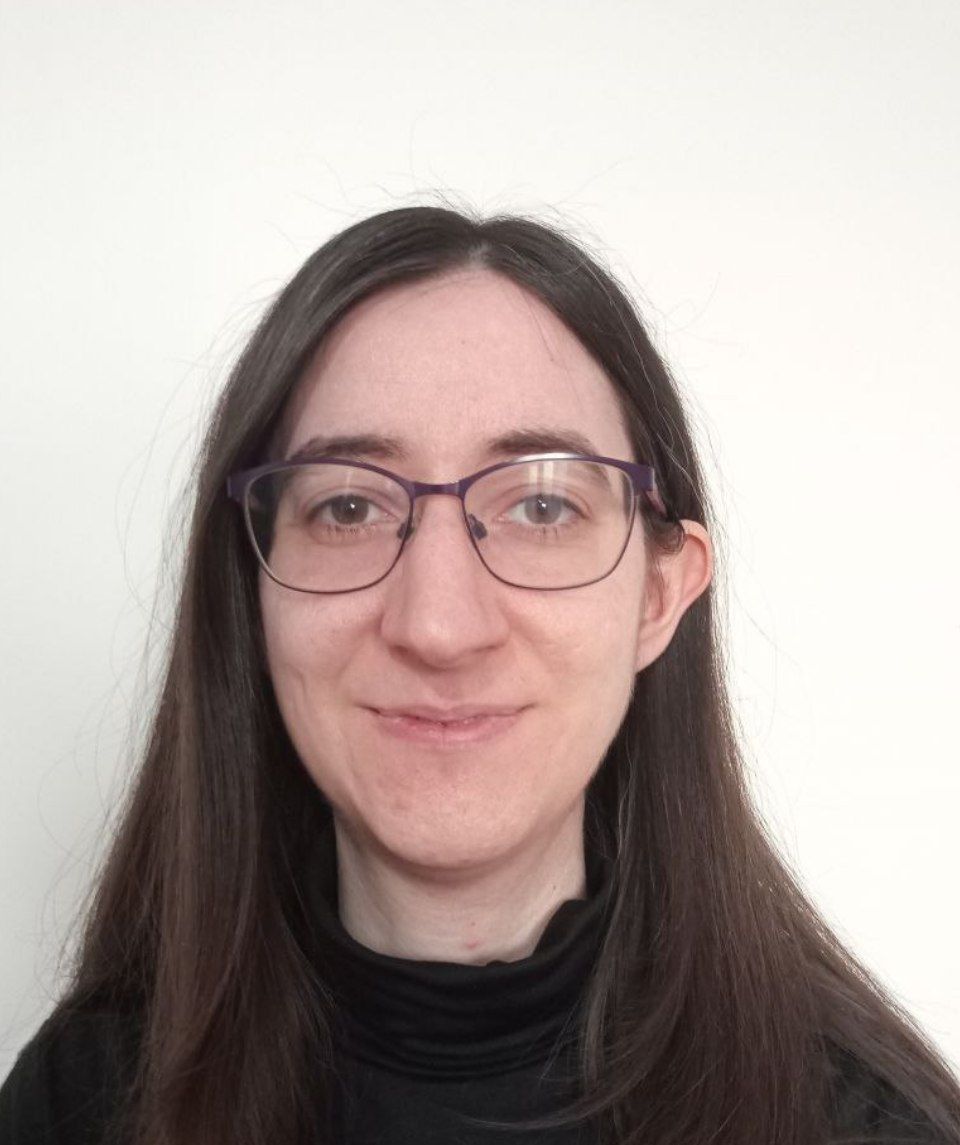}}]{Stefania Cucco} received Bachelor’s degree in Physical Engineering and Master’s degree in Electronic Engineering (curriculum “Micro and Nano Systems”) at Politecnico di Torino, Italy, in 2017 and 2022 respectively. On November 2022 she started the PhD in Electrical, Electronics and Communications Engineering (EECE) at same Politecnico. 
Her PhD research continue the master thesis project relative to ring resonator design for integrated hybrid laser in nonlinear model of microring resonator with complex waveguide cross-section, stable and narrow linewidth.
\end{IEEEbiography}
\vskip 0pt plus -1fil
\begin{IEEEbiography}[{\includegraphics[width=1in,height=1.25in,clip,keepaspectratio]{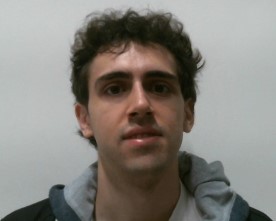}}]{Marco Novarese} received the BSc degree in Physical Engineering  and the MSc degree in
Nanotechnologies For ICTs both  from Politecnico di Torino, Turin, Italy, in 2017 and 2019 respectively.  
He obtained his PhD diploma at Politecnico di Torino in 2023 with a thesis on  modelling and characterisation of 
microrings for semiconductor lasers integrated in the Silicon Photonics platform.
Since Septemember 2023, he is Post-Doc at Politecnico di Torino. His current research is focused on the development
of an experimental setup for the analysis of VCSELs and silicon photonic hybrid laser. 
Other reasearch interests include stress measurements on quantum dots laser, modeling of free carrier diffusion in silicon and polysilicon ring resonators, and nonlinear laser dynamics.
\end{IEEEbiography}
\vskip 0pt plus -1fil
\begin{IEEEbiography}[{\includegraphics[width=1in,height=1.25in,clip,keepaspectratio]{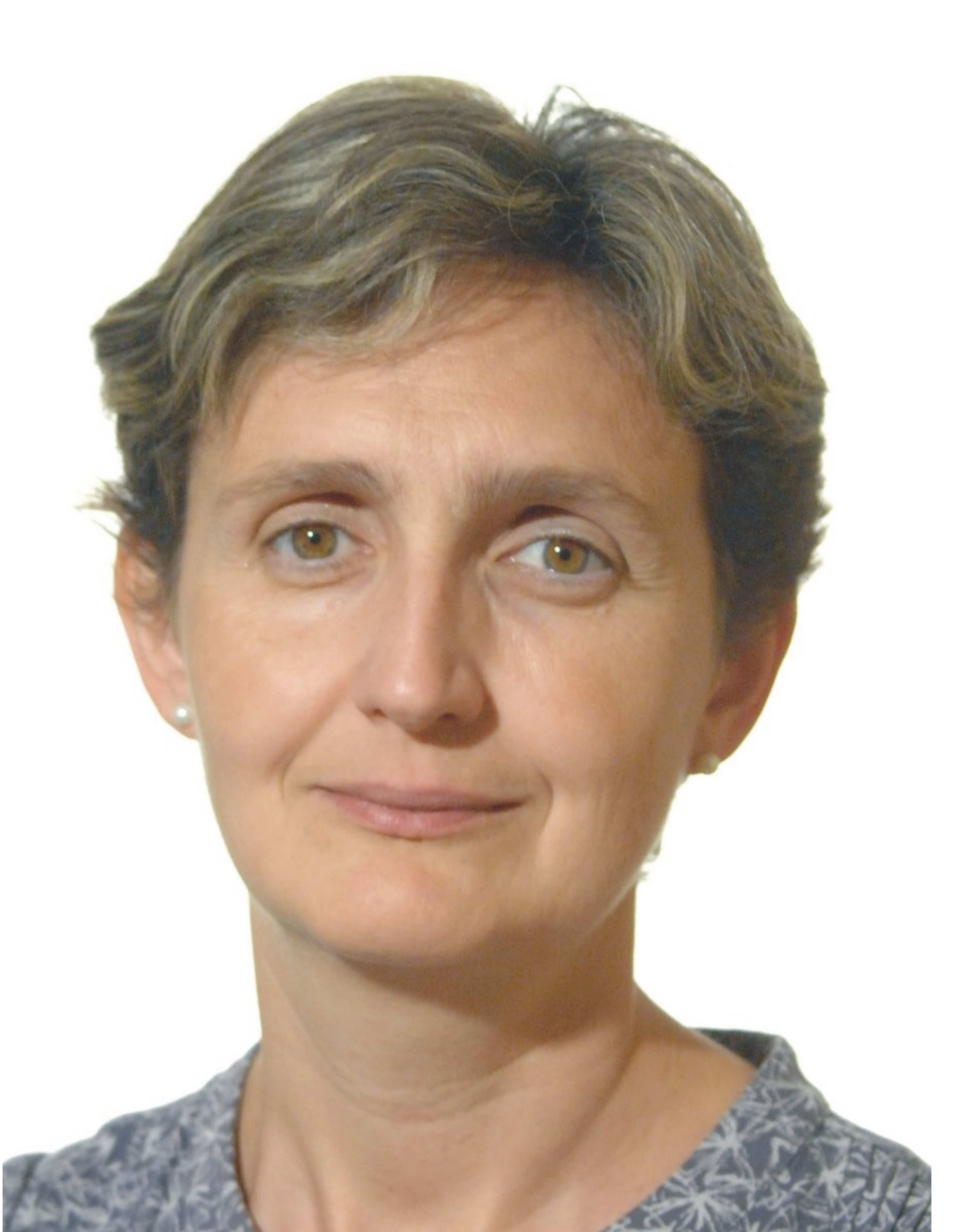}}]{Mariangela Gioannini} received the MS degree in Electronic Engineering and the PhD degree Electronics and Communication Engineering both from Politecnico di Torino in 1998 and 2002 respectively. Since 2005 she is with Dipartimento di Elettronica e Telecomunicazioni of Politecnico di Torino where she is now full professor. She has carried on research on the numerical modelling of semiconductor lasers and optical amplifiers based on III-V materials including quantum dot lasers and QCLs. In recent years research activity has been extended to silicon photonics and VCSELs including the coordination of an experimental lab dedicated to VCSELs and silicon photonics PICs. Since 2013 she has been in the technical program committee of major conferences in the area of photonics: CLEO-Europe, CLEO US, IEEE ISLC and Photonics West. Since 2022 she is Associate Editor of IEEE Photonic Technology Letters.

\end{IEEEbiography}

\vspace{11pt}

\vfill

\end{document}